%% file: bsmith.tex
\documentclass[preprint2]{aastex}

\def\sun{\hbox{$\odot$}}
\def\farcm{\hbox{$.\mkern-4mu^\prime$}}
\def\farcs{\hbox{$.\!\!^{\prime\prime}$}}

\defcitealias{smi05}{Paper I}

\shorttitle{Knots of Star Formation}
\shortauthors{Smith et al.}

\begin{document}

\title{A Comparative Study of Knots of Star Formation in Interacting vs.\ Spiral Galaxies}

\author{
Beverly J. Smith\altaffilmark{1}, 
Javier Zaragoza-Cardiel\altaffilmark{2},
Curtis Struck\altaffilmark{3},
Susan Olmsted\altaffilmark{1},
Keith Jones\altaffilmark{1}
}

\altaffiltext{1}{Department of
Physics and Astronomy, East Tennessee
State University, Johnson City TN  37614;
Southeastern Association for Research in Astronomy;
smithbj@etsu.edu}
\altaffiltext{2}{Instituto de Astrofisica de Canarias,
La Laguna, Tenerife, Spain; 
Now at 
Instituto de Astronom\'ia, Universidad Nacional Aut\'onoma de M\'exico, 
Mexico
City, Mexico
}
\altaffiltext{3}{Department of Physics and Astronomy, Iowa State University, Ames IA  50011}

\begin{abstract}

Interacting galaxies are known to have higher global rates of star formation
on average
than normal galaxies, relative to their stellar masses.
Using UV and IR photometry combined with new and published H$\alpha$ images,
we have compared the star formation rates of 
$\sim$700
star forming complexes in 46 nearby interacting 
galaxy pairs with those of regions in 39 normal spiral galaxies.
The interacting galaxies have
proportionally more regions with high star formation rates 
than the spirals.
The most extreme regions in the interacting 
systems lie at the intersections of spiral/tidal structures,
where gas is expected to pile up and trigger star formation. 
Published Hubble Telescope images show unusually large and luminous
star clusters in the highest luminosity regions.
The star formation rates of the clumps correlate with measures
of the dust attenuation, consistent with the idea that regions with more
interstellar
gas have more star formation.
For the clumps with the highest star formation rates, the 
apparent dust attenuation is consistent with the Calzetti
starburst dust attenuation law.  
This suggests that
the high luminosity regions are dominated by a central group of 
young stars 
surrounded by a shell of clumpy
interstellar gas.  In contrast, the lower luminosity clumps are 
bright in the UV relative to H$\alpha$, suggesting either 
a high differential attenuation between the ionized gas and the 
stars,
or a post-starburst population bright in the UV but faded in H$\alpha$.
The fraction of the global light of the galaxies in the clumps
is higher on average for the interacting galaxies than for the spirals.
Thus 
the star forming regions in interacting galaxies are 
more luminous, dustier, or younger on average.

\end{abstract}


\section{Introduction}

Since the pioneering study of \citet{larson78}, 
numerous studies have concluded that gravitational interactions
between galaxies can trigger star formation 
\citep{lonsdale84, keel85, bushouse87, bushouse88, kennicutt87, 
barton00, barton03, lambas03, nikolic04, woods06, smith07, 
lin07, 
ellison08,
li08, woods10}.
Observations in 
the 
H$\alpha$ line \citep{bushouse87, kennicutt87},
the 
far-infrared \citep{bushouse87, bushouse88, kennicutt87},
and 
the 
mid-infrared \citep{lonsdale84, smith07, lin07} along
with optical spectroscopic studies \citep{li08} indicate that 
the 
global star formation rates (SFRs)
of strongly
interacting pre-merger galaxy pairs 
are enhanced by 
about a factor of 2 $-$ 3 on average 
relative to their stellar masses
compared to normal galaxies, however, there is significant scatter
from galaxy to galaxy.
On average, closer pairs tend to have higher rates of star formation 
\citep{barton00, barton03, lambas03, nikolic04, woods06, woods10, lin07, li08}.
The star formation in interacting
galaxies is often centrally-concentrated \citep{lonsdale84, keel85, 
bushouse87,
smith07},
likely 
due to angular momentum transfer
driving gas into the inner core (e.g., \citealp{barnes91},
\citealp{mihos96},
\citealp{dimatteo07}).

In addition to enhanced nuclear star formation,
strong 
star formation is sometimes seen in the outer disks and tidal structures
of interacting galaxies 
\citep{schweizer78, mirabel91, mirabel92, hibbard96,
smith10}.
The 
tidal structures 
of interacting
galaxies frequently display
star formation
morphologies that usually are not seen in isolated galaxies. 
These include: 1) regularly-spaced star formation regions (`beads on a string') 
along tidal features and spiral arms 
\citep{hancock07, smith10}, 
2) massive concentrations of stars and gas near the ends of tidal features, the so-called tidal dwarf galaxies (TDGs) (e.g., 
\citealp{duc94, duc97, duc00, smith10}),
3) luminous star forming regions at the base of tidal features 
(`hinge clumps', 
\citealp{hancock09, smith10, smith14}), 
and 4) gas-rich star-forming structures 
produced by accretion from one galaxy to another (e.g., 
\citealp{smith08}).
Beads-on-a-string may be indicative of the 
accumulation scale of local gravitational 
instabilities
(e.g., \citealp{elmegreen96}),
while TDGs and some accretion star formation may 
result from gas pile-ups and subsequent gravitational
collapse \citep{duc04, wetzstein07, smith08}.
Hinge clumps may be produced by
converging flows of dissipative gas
along caustics, where a caustic is a narrow pile-up zone 
caused by orbit crowding
\citep{struck12, smith14}.
Hinge clumps, accretion knots, and TDGs, although rare in the local
Universe, were likely much more common in the past.   
Hinge clumps
bear an intriguing resemblance to the massive star forming
clumps seen in high redshift disks (e.g.,
\citealp{elmegreen09, forsterschreiber11}), while
the prolonged infall out of a tidal tail may be the best local analog 
to cold accretion onto galaxies, evidently a common process in the early 
Universe, but not in the present. 

Theoretical studies suggest that gas turbulence 
and pressure
are enhanced in galaxy mergers, potentially leading to larger Jeans masses,
more massive star forming regions,
larger fractions of dense gas,
and more efficient star formation
\citep{elmegreen93, bournaud08, teyssier10,
bournaud11, renaud12, powell13, renaud14}.
Consistent with this scenario, larger gas velocity
dispersions have been found in
some interacting galaxies 
\citep{elmegreen93, irwin94, elmegreen95, zaragoza13, zaragoza14, zaragoza15}.
Based on the observed velocity dispersion measurements, 
\citet{zaragoza13, zaragoza14, zaragoza15}
have concluded that the highest mass star forming regions in interacting
galaxies are gravitationally-bound, while lower mass
regions are pressure-confined.   This suggests a different mode
of star formation may be operating in the highest luminosity
regions.

In spiral galaxies, the global SFR depends upon the gas surface density
via a power law relation 
\citep{schmidt59, kennicutt89b,
kennicutt98}.
Within the disks of nearby spirals,
a similar relation is found on smaller spatial scales 
\citep{kennicutt07, liu11}.
However, 
extreme starburst galaxies (many
of which are interacting or merging) may have 
enhanced star formation relative to the gas content
compared to the relationship for spirals
\citep{daddi10, genzel10, saintonge12, silverman15},
suggesting the existence of 
a distinct starburst `mode' of star formation. 
Such a mode had been predicted earlier on the basis of theoretical
arguments, with the starburst phase being limited by stellar feedback
\citep{scalo86, quillen08}.
High resolution simulations suggest that increased gas turbulence,
cloud filamentation, and cloud fragmentation 
in interacting galaxies may lead to 
more efficient star formation \citep{teyssier10, renaud14}.
However, the existence of two distinct 
modes of star formation is uncertain
\citep{kennicutt98, powell13}.

In addition to having higher SFRs, 
interacting galaxies 
may have more obscured young stars on average.
Interacting galaxies have larger far-infrared-to-H$\alpha$
luminosities on average than isolated systems
\citep{bushouse87}, suggesting higher H$\alpha$ extinctions
in the interacting systems.
The global optical and UV colors of interacting
galaxies show a larger dispersion than those of normal systems
\citep{larson78, solalonso06, smith10b},
likely
a consequence of both more young stars and more dust attenuation.
Global UV-to-MIR ratios also imply larger 
attenuations in galaxy
pairs than in normal galaxies \citep{yuan12}.
As gas gets driven into the central regions of galaxies
by an interaction,
not only is the SFR increased
but
also the dust column density.

More generally, the dust attenuation in galaxies tends to increase
with increasing SFR 
\citep{wang96, buat98, buat99, hopkins01, sullivan01}.
The galaxies with the highest far-infrared luminosities (L$_{\rm FIR}$)
are 
generally interacting or merging systems
\citep{smith87, armus87, sanders88, melnick90,
leech94, clements96, clements96b, sanders96, rigopoulou99}.
The highest L$_{\rm FIR}$ galaxies
have large
far-infrared-to-optical
ratios 
\citep{smith87, armus87} implying high dust attenuation.

How much the starlight in a galaxy is attenuated at each wavelength
depends upon the amount of dust, the type of dust,
and the distribution of the dust relative to the stars.
Differences in the assumed geometry of the system can cause large
variations in the implied dust attenuation law \citep{natta84, witt00, charlot00,
granato00, wild11}.
Contributing
factors include the clumpiness of the gas, the fraction of diffuse dust vs.\ 
dust associated with the birth cloud,
and the relative amounts of attenuation
of stars of various ages. 
There is evidence that the dust attenuation law 
in normal spirals may differ from that in starbursts 
\citep{burgarella05, panuzzo07, 
boquien09, boquien12, 
conroy10,
mao14}.  The assumed corrections for dust attenuation can make
large differences in the inferred SFRs, luminosities,
and stellar masses of galaxies.  
At present, there is considerable uncertainty
in the attenuation law applicable for different
types of galaxies.

To better understand star formation triggering mechanisms
and dust attenuation in interacting galaxies,
spatially-resolved studies of nearby galaxies are crucial.
By combining 
multi-wavelength 
observations 
of individual
star forming regions in the galaxies with 
numerical simulations of
the interaction, we can better identify what is triggering
the star formation in that particular system.
Over the last several years, we have
conducted detailed multi-wavelength studies
of star-forming `clumps' within five individual galaxy pairs
using aperture diameters of 0.8 $-$ 6.2 kpc,
and have constructed
matching numerical models of the interactions:
Arp 82 \citep{hancock07}, 
Arp 107 \citep{smith05, lapham13},
Arp 284 \citep{struck03, smith05b, peterson09},
Arp 285 \citep{smith08},
and Arp 305 \citep{hancock09}.
Other researchers have 
done similar studies of regions in other
interacting galaxies (e.g.,
Arp 24: \citealp{cao07};
Arp 85: \citealp{calzetti05, calzetti07};
Arp 158: \citealp{boquien11};
Arp 244: \citealp{zhang10};
NGC 2207/IC 2163: \citealp{elmegreen95, elmegreen01, 
elmegreen06, 
struck05, kaufman12}).
These `clumps' are star formation complexes 
containing multiple star clusters (e.g., \citealp{elmegreen06,
peterson09, smith14}).

Such studies of individual systems provide
clues to star formation triggering 
mechanisms
on a case-by-case basis, however, they do not give
much information about how important statistically the
different processes are to galaxy evolution as a whole.
For a more comprehensive understanding of star formation
and dust attenuation
in interacting galaxies, surveys of regions within 
multiple galaxies would be valuable.   
A few such comparative studies have been done recently
\citep{boquien09, boquien10, lapham13, smith14}, but they 
only involve a handful of galaxies.

In the current study, we 
describe 
a systematic multi-wavelength
investigation of the star forming regions within a sample of 46
interacting galaxy pairs, and compare with a matching set
of normal spiral galaxies.
This is a much larger number of galaxies than included in earlier
such surveys.
We have extracted UV/optical/IR photometry for star forming regions
within these galaxies, and have estimated SFRs and dust attenuations 
for these regions.
In Section 2 of this paper, our samples of galaxies are discussed in detail.
Our datasets are described in Section 3.   For this study,
we acquired new narrowband H$\alpha$ images of some of the
systems; these observations are described in Section 3.
The clump selection and photometry is discussed in Section 4.
The star formation rates for the regions are discussed in Section 5.
In Section 6, we discuss the dust attenuation in the clumps.
Our results are summarized in Section 7.

\section{Galaxy Samples }

In this study, 
we compare the properties of star forming
regions within interacting galaxies
with those in normal spiral galaxies.   
The selection of these samples is
described below.

\subsection{Pre-Merger 
Interacting Galaxies with Strong Tails and Bridges}

For the interacting galaxies, we use
our `Spirals, Bridges, and Tails' (SB\&T) sample
\citep{smith07, smith10}.
This consists of 
more than three dozen
pre-merger
galaxy pairs chosen
from the \citet{arp66} Atlas of Peculiar Galaxies 
to be 
relatively isolated binary systems with strong tidal distortions;
we eliminated merger remnants, close triples, and multiple systems.
They have radial velocities $<$10,350 km~s$^{-1}$ and angular sizes
$\ge$3$'$.  
The interacting
pair NGC 4567/8 was included in the SB\&T sample in \citet{smith07}, 
as it fits this basic
selection criteria although it is not in the Arp Atlas.
To the original SB\&T sample, 
we have added a few additional
Arp systems with 
slightly smaller angular sizes 
that were not in
our earlier studies.
We have also added the similar system NGC 2207/IC 2163 
\citep{elmegreen95, elmegreen01, elmegreen06}, which
is not in the Arp Atlas.

Only galaxies with Spitzer 8 $\mu$m maps covering
both galaxies in the pair were included in the final
sample for the
current study,
since we use the 8 $\mu$m images to select the target star forming
regions (see Section 4).
Our final sample of interacting galaxies contains 46
pairs\footnote{Arp 297 consists of two pairs at different
redshifts, Arp 297S and Arp 297N.  These two pairs are treated
separately in this analysis.}.
This sample is given in Table 1, along with the distance to each
system, the far-infrared luminosity, the total H$\alpha$ luminosity,
and 
the reference for the H$\alpha$ luminosity.
Throughout this paper we use 
distances from the NASA Extragalactic
Database (NED\footnote{http://ned.ipac.caltech.edu}), 
assuming
H$_0$ = 73 km~s$^{-1}$~Mpc$^{-1}$ and accounting for peculiar velocities
due to the Virgo
Cluster, the Great Attractor, and the Shapley Supercluster. 
A histogram of the distances to the sample galaxies is provided
in the top panel of Figure 1.

The SB\&T galaxies are relatively simple systems compared
to many interacting galaxies, for example, galaxies in compact groups
or more advanced mergers, thus they
are more 
suitable for the detailed matching of simulations
to observations.
By excluding advanced mergers from our sample, we are 
omitting some of the most extreme starbursts in the 
local Universe, which are sometimes later-stage mergers than our
sample galaxies (e.g., \citealp{melnick90, rigopoulou99}).
The galaxies in our interacting sample contain the closest
and best-studied
examples of strong tidal tails and bridges in the local Universe, with
a median distance of only 48 Mpc.  This is much closer than other samples
selected from other surveys, for example, 
the 
Sloan Digitized Sky Survey (SDSS) selected
TDGs
studied by \citet{kaviraj12}
have a median distance of 
220 Mpc, 
while the 
\citet{torres09}
compact groups containing TDGs 
have a median distance of 
62 Mpc.
The SB\&T sample contains about a dozen examples of `beads
on a string' and about half a dozen `hinge clumps'.  It also contains
11 candidate TDGs, and about half a dozen extended
structures likely caused by mass transfer between galaxies
\citep{smith10}.

\subsection{A `Control' Sample of Spiral Galaxies}

In \citet{smith07},
we compared the distribution of Spitzer broadband mid-infrared
colors for the SB\&T galaxies 
with those of a sample of 26 `normal' spirals
selected from the SINGS sample \citep{kennicutt03},
after
eliminating SINGS galaxies with massive nearby companions
(i.e., velocity difference
$\le$1000 km~s$^{-1}$, optical luminosity
$>$ 1/10 the target galaxy, and separation from the target
galaxy of $<$ 10 times the diameter of target galaxy or the 
companion, whichever is larger).
In \citet{smith10b}, we constructed an alternative control sample
of spirals. 
We started with the 
`GALEX Ultraviolet Atlas of Nearby Galaxies'
\citep{gildepaz07}, 
and 
selected the subset 
with SDSS images that are 
classified as normal spirals and have distances $<$ 143 Mpc.  
NED
and available GALEX/SDSS images
are used to eliminate galaxies with nearby massive companions, leaving a total
of 121 spiral galaxies.
We compared the large-scale environments and blue luminosity
distribution of this sample with those of the SB\&T sample, and
found they are statistically indistinguishable 
\citep{smith10b}.

In the current study, our final 
control sample of spirals (Table 2) consists of all 39 of the
galaxies in the union of these two spiral samples 
that have Spitzer 8 $\mu$m images available.
A histogram of the distances to the sample spiral galaxies is provided
in the bottom panel of Figure 1.
The median distance to the spiral galaxies is 14 Mpc, thus the spirals
are on average closer than the interacting systems.
Later in this paper, we investigate 
the subset of galaxies
closer than 67 Mpc separately (see Section 4.1). 
All but two of the spirals are within 67 Mpc, while 16 out
of the 46 interacting pairs have distances greater than 67 Mpc.

In the top panel of
Figure 2, we provide a histogram of the global Spitzer 3.6 $\mu$m
luminosities ($\nu$L$_{\nu}$) of the individual galaxies in
the interacting galaxy pairs.
The total 3.6 $\mu$m fluxes were obtained from
Spitzer images (Section 3) as in \citet{smith07}.
The lower panel of Figure 2 shows a similar histogram for the
spiral galaxies.  
The global Spitzer 3.6 $\mu$m luminosities L$_{3.6}$ of galaxies
is an approximate measure of
their stellar mass, as it is usually dominated by
the light from older stars (e.g., \citealp{helou04}).
Figure 2 shows that the spirals and the interacting galaxies
have similar distributions of L$_{3.6}$,
thus there are not large differences
between the stellar masses of the galaxies in the two samples
on average.
Proportionally, there are a few more lower 
luminosity galaxies in the interacting
sample, likely because of the inclusion of low mass companion
galaxies.  
The interacting galaxy sample also contains a few more higher
luminosity galaxies than the spiral sample.
The subset of interacting galaxies with distances less
than 67 Mpc has a L$_{3.6}$ distribution very similar to that
of the spiral sample.

\section{Datasets}

In this study, we use 
GALEX near-UV (NUV) 
and far-UV (FUV) maps, along with 
Spitzer near-infrared
(3.6 $\mu$m and 4.5 $\mu$m) and 
mid-infrared (5.8 $\mu$m, 8.0 $\mu$m, and 24 $\mu$m)
images.
We also include 
new and archived H$\alpha$ images in this study.
An Atlas of the Spitzer infrared images
of the SB\&T galaxies
was published in 
\citet{smith07}.
A second Atlas of the SB\&T galaxies
displaying the GALEX and corresponding SDSS images
was published in
\citet{smith10}. The global UV, optical, and IR properties of
the SB\&T galaxies were
compared with those of normal spiral galaxies in
\citet{smith07, smith10} and \citet{smith10b}.

The 3.6 $\mu$m $-$ 8.0 $\mu$m images
were acquired using the Spitzer Infrared Array Camera
(IRAC; \citealp{fazio04}), while the 24 $\mu$m images were
obtained with the Spitzer Multiband Imaging Photometer
for Spitzer (MIPS; \citealp{rieke04}).
The Spitzer images used in the current study
were acquired from the Spitzer Heritage
Archive\footnote{http://sha.ipac.caltech.edu}, except for 
four large angular size SINGS galaxies, for which larger mosaicked images
were obtained from the NED image database.  
The 24 $\mu$m images have 2\farcs45~pixel$^{-1}$.
The 3.6 $\mu$m $-$ 8 $\mu$m images from the Heritage
archives have 0\farcs6~pixel$^{-1}$; the SINGS IRAC
images initially had 0\farcs75~pixel$^{-1}$, which we rebinned
to match the other images.
The Spitzer FWHM spatial resolution is
1\farcs5 $-$ 2$''$ for
the 3.6 $\mu$m $-$ 8 $\mu$m bands, 
and $\sim$6$''$ at 24 $\mu$m. 
For more details on the Spitzer observations
and the global fluxes,
see \citet{smith07}.

We only used GALEX images with exposure times greater than 800 sec.
The GALEX FUV band has an effective wavelength
of 1516 \AA~ with a full width half maximum (FWHM) of 269 \AA, 
while the NUV band has an effective
wavelength of 2267 \AA ~and FWHM 616~\AA.
The GALEX images have 1\farcs5 pixels, and the point spread
function has a FWHM of $\sim$ 5$''$.
For more details about the GALEX 
observations and the global fluxes, see \citet{smith10}.

H$\alpha$ maps are available for the majority of
the galaxies in our sample, either from our own
observations or from published archives.
We acquired new H$\alpha$ optical images for some of
the galaxies using the 0.9m optical 
telescope of the Southeastern Association for Research in Astronomy
(SARA) on Kitt Peak in Arizona\footnote{The SARA 0.9m 
telescope at Kitt Peak is owned and 
operated by the Southeastern Association for Research in Astronomy 
(saraobservatory.org).}
or the 4.2m William Herschel Telescope
(WHT) at the Roque de los Muchachos Observatory, La Palma (Spain).
These galaxies are identified in Table 1 and 2.
For the SARA telescope, we used an Axiom/Apogee 2048 $\times$ 2048 CCD
with binning set to 2 $\times$ 2 resulting in a pixel size of 
0\farcs51~pixel$^{-1}$.
For the WHT we used the ACAM (Auxiliary-port Camera; 
\citealt{benn08}), an
instrument mounted permanently at the WHT for broad-band and
narrow-band imaging. The observations with ACAM resulted in a pixel size of
0.25$''$~pixel$^{-1}$.
We obtained images of these galaxies in narrowband 
(FWHM = 14 $-$ 50 \AA)
redshifted
H$\alpha$ filters matched to the redshift of the galaxy and a
second narrowband image off of the H$\alpha$ line.
In a few cases, a broadband red filter consistent with the SDSS
r filter was used instead for the continuum.
The seeing was typically 0\farcs7 $-$ 1\farcs5.
Continuum subtraction was accomplished as in \citet{gutierrez11}.
Spectrophotometric standard stars were also observed when the sky was clear.

For most of the remaining galaxies in our sample, 
published continuum-subtracted
H$\alpha$ maps are 
available from NED (see Tables 1 and 2 for references).
In some cases, matching narrowband
off-H$\alpha$ maps or broadband R images are available from the same
source. 
When necessary, we registered these H$\alpha$ and continuum
maps to match the Spitzer
images.
We used published calibration information for these images when available;
otherwise, we calibrated the images using
published total H$\alpha$ fluxes from 
\citet{kennicutt09}.
When necessary, the H$\alpha$ fluxes have been approximately
corrected for the nearby [N~II] lines in the filter.
We assume a typical uncertainty on the H$\alpha$ fluxes of $\sim$30$\%$,
but this likely varies from galaxy to galaxy.
The total H$\alpha$ luminosity for each system is provided in Tables 1
and 2.

\section{Clump Selection and Photometry}

\subsection{Selection }

Our goal in this project is to compare the properties of
star formation regions
in a sample of interacting galaxies with those in normal spirals.
We want to study the same physical scale within each galaxy.
As a compromise between our desire for detailed spatially-resolved
studies of these galaxies, the limiting resolution 
of the GALEX and Spitzer 24 $\mu$m images, 
and our desire to have the largest sample possible, we have investigated
star formation on two different physical scales within the galaxies.

First, for the entire sample of galaxies, we have selected
clumps and extracted photometry
using
an aperture radius of 2.5 kpc.
This scale is set by
the limiting aperture radius of $\sim$3\farcs0 for accurate
photometry on the GALEX and Spitzer 24 $\mu$m images
(see Section 4.3).
For the most distant galaxy in our sample, Arp 107, at 142 Mpc,
2.5 kpc corresponds to
3\farcs6.
A radius of 2.5 kpc 
corresponds to 42\farcs5 at the distance of the closest interacting
galaxy in our sample, Arp 85 (M51), 12.1 Mpc.
The 
closest of the spiral galaxies, NGC 7793, is at 3.28 Mpc,
giving an aperture radius of 157\farcs2.

Second, to investigate star formation on smaller scales, for a subset
of galaxies we have also conducted clump selection
and photometry using an aperture radius
of 1.0 kpc.   
With a limiting aperture radius of 
3\farcs0,
this subset
of galaxies is limited to systems within 67 Mpc.   
The 1.0 kpc 
subset consists of 30 interacting pairs and 37 spiral galaxies.
The aperture radii used for the 1.0 kpc sample range from 3\farcs1 to 
24\farcs3.

As in our earlier studies of Arp 107, Arp 82, and Arp 285 
\citep{smith05, hancock07, smith08, lapham13}, 
we selected our sample of
clumps using the Spitzer
8 $\mu$m images.  
Although the 24 $\mu$m bandpass and the H$\alpha$ filter are
typically considered better
tracers of star formation than 8 $\mu$m
(e.g., \citealp{calzetti05, calzetti07}),
the 24 $\mu$m images have 
lower spatial resolution than the 8 $\mu$m Spitzer
images and 
are more likely to have artifacts, while the H$\alpha$ dataset is
inhomogeneous. 
Further, a few of the galaxies in our sample lack 24 $\mu$m and
H$\alpha$ images.
Since older stars may contribute 
to powering the 8 $\mu$m emission
from star forming regions
in some cases
\citep{calzetti05, calzetti07}, some of the clumps we select may 
have older ages than regions selected based on 24 $\mu$m
or H$\alpha$.  
However, more than half
of
the 8 $\mu$m-selected clumps 
in our earlier studies
have best-fit UV/optical population synthesis ages 
less than 10 Myrs \citep{hancock07,
smith08, lapham13}, thus we are not
strongly biased against young regions.

To select clumps over a consistent spatial
scale from galaxy to galaxy, we first smoothed the 
8 $\mu$m images using a Gaussian to produce a final full width
half maximum (FWHM) resolution of 2.5 kpc or 1.0 kpc, 
half our selected aperture
diameter.
For the most distant galaxies in the
sample, no smoothing is needed, as this resolution
is equal to the native resolution of the 8 $\mu$m image.

Clumps were then selected automatically from the smoothed images using 
the {\it daofind}
routine 
\citep{stetson87} 
in the 
Image Reduction and Analysis Facility
(IRAF\footnote{http:$//$iraf.noao.edu})  
software.   With {\it daofind}, we 
used a detection threshold
of 10$\sigma$ above the noise
level in the smoothed images, and 
set the {\it sharplo, sharphi, roundlo,} and {\it roundhi}
parameters to 0.1, 1.2, $-$2.0, and 2.0, respectively, to allow
slightly extended and/or elongated clumps.
We visually inspected each selected source by eye, to eliminate
spurious detections due to artifacts in the images.

We note that our aperture sizes are significantly larger than
typical sizes of individual H~II regions.  For comparison, the diameter
of the Orion
nebula is $\sim$5 pc, while the 30 Doradus
region in the Large Magellanic Cloud has a diameter $\sim$ 400 pc,
and the giant H~II regions NGC 5471 in M101 
and NGC 604 in M33 have diameters of $\sim$ 800 pc and $\sim$ 400 pc,
respectively \citep{kennicutt84}.
Thus our `clumps' are likely complexes of multiple H~II regions.

\subsection{Clump Classification}

We classified the clumps in the interacting
galaxies into three basic groups:
clumps in the inner disks of the galaxies, clumps in the tidal features,
and galactic nuclei.  Clumps in the spiral galaxies were classified
as either nuclei or disk clumps.
We used the Spitzer 3.6 $\mu$m images to 
distinguish galactic nuclei from other clumps.
The classification of clumps as `tidal' vs.\ `disk' is somewhat subjective.
As in \citet{smith12}, 
when the source falls inside SDSS g (4680 \AA) contours that have a 
smoothly elliptical shape, we classified it as a disk source.
Alternatively, if considerable asymmetry is visible in the surrounding
SDSS contours,
the source is classified as tidal.
Some example galaxies with their clumps marked are shown in Figure 3.

In addition to clumps associated with
the galaxies, we also identified 8 $\mu$m point sources
outside of the galaxies, which we classify as `off' sources.
Some of these 
`off' sources may be detached tidally-formed star forming regions
associated with the galaxies;
alternatively, they may be background or foreground objects.
We use these `off' sources as a comparison sample for the tidal and
disk objects (see Appendix).
As a dividing boundary between objects classified as `off' sources and 
`disk'/`tidal' sources
assumed to be associated with the galaxies, for the galaxies with
SDSS images available we use
an SDSS g filter 
isophote of 24.58 mag/arcsec$^2$.  This is approximately
equivalent to B = 25 mag/arcsec$^2$, using the median g $-$ r
color for tidal tails from \citet{smith10} and the \citet{jester05}
g to B conversion.

For systems with no SDSS images available, we use a GALEX NUV
surface brightness of 26.99 mag/arcsec$^2$, applying the median
NUV $-$ g color of tidal tails of 2.4 \citep{smith10}.
For the three systems with neither SDSS nor GALEX images, 
we used a Spitzer 3.6 $\mu$m isophote of 20.91 mag/arcsec$^2$,
using the median g $-$ [3.6] color of 3.67 for tidal tails 
\citep{smith07, smith10}.
Some of the sources we list as `disk' and `tidal' may
not be physically associated with the galaxies,
but instead may be foreground or background objects.  This topic is
discussed further in Section 4.5 and the Appendix.

In 
the interacting galaxy 1.0 kpc clump sample, there are
514 disk clumps, 208 tidal clumps, and 162 `off' clumps, in addition to
the 60 nuclear regions.  In the 1.0 kpc radius clump sample for the spiral
galaxies, there are 997 disk clumps and 85 `off' regions, as well
as 37 nuclear regions.  
The 2.5 kpc clump sample for the interacting galaxies has 151 disk
clumps, 252 tidal regions, and 190 `off' sources, along with 92 nuclear
sources.   There are 159 disk clumps in the 2.5 kpc radius clump sample
for the spiral galaxies, along with 51 `off' sources and 39 nuclei.

\subsection{Photometry}

From the Spitzer, GALEX, and H$\alpha$ images,
we extracted aperture photometry of the 
clumps from the unsmoothed images
using the 
IRAF
{\it phot} routine.
As noted above, for this photometry we use radii of either 2.5 kpc
or 1.0 kpc, depending upon the sample.
We used a sky annulus with the mode sky
fitting algorithm, with an inner
radius equal to the aperture radius, and an annulus width
equal to 1.2 $\times$ the aperture radius.
The mode option for determining the background
is useful in crowded fields.

For the more distant galaxies in the sample, 
aperture corrections
are needed to account for flux spillage outside of the  
aperture.   
This is particularly important 
for the GALEX and Spitzer 24 $\mu$m images, which are relatively low 
spatial resolution; however, we also utilize aperture corrections
for the Spitzer IRAC images.
For the H$\alpha$ photometry, we did not apply
aperture corrections since
these corrections are expected to be small.
For the Spitzer images, we used aperture corrections from
the IRAC and MIPS Instrument 
Handbooks\footnote{http://irsa.ipac.caltech.edu/data/SPITZER/docs/},
interpolating between
the tabulated values.
For the GALEX images,
we calculated aperture corrections for 
each image individually, by 
doing aperture photometry for three to ten moderately bright isolated
point sources in the field.
For these stars, we extracted the fluxes within our target apertures,
and compared with that obtained using a 17$''$ radius.
No aperture corrections were used for 
GALEX for 
aperture radii 
$\ge$17$''$.
If the clumps are somewhat resolved at these spatial scales,
then aperture corrections based on stars in the field
will be under-estimates.
The aperture corrections 
for the GALEX bands 
are plotted against aperture radius in Figure 4.
Note that the NUV aperture corrections are more consistent from image
to image than those at FUV.

We corrected the UV photometry for Galactic reddening by starting with
optical estimates from 
\citet{schlafly11} as provided by NED, and extrapolating to the UV
using the \citet{cardelli89} attenuation law.

\subsection{Fraction Global Light in Clumps}

For each galaxy in the sample, we co-added the fluxes for the clumps in that 
galaxy, and compared the combined light of the clumps 
with the global
fluxes of the galaxies as tabulated in \citet{smith07, smith10}
and \citet{smith10b}.
The fraction of the total light from the galaxy due
to the targeted clumps varies
from galaxy to galaxy and wavelength to wavelength.
On average for all wavelengths, the fraction of the total light in
clumps is about 20\% higher for the 2.5 kpc clump sample than for
the 1.0 kpc clump sample.
At 8 $\mu$m for the 1.0 kpc sample, the median
fraction in clumps is 27\% for the spirals and 34\% for the 
interacting galaxies.  
For the 2.5 kpc sample, the median
fraction in clumps at 8 $\mu$m is 38\% for the spirals and 58\% for the 
interacting galaxies.
Within these samples, there is a large scatter 
from galaxy to galaxy, likely depending upon the sensitivity of the
image, the morphology of the galaxy, and the age of the regions.
Similar fractions are found for the FUV and H$\alpha$ images, while
the NUV images have smaller fractions on average (median
fractions in clumps between 17\% and 42\%, depending upon the sample). 
At 24 $\mu$m, the
median fraction in clumps is about 60\% higher on average than 
at 8 $\mu$m.
These results are consistent with a picture in which the 
NUV light originates from older more dispersed stars than the stars
traced in the other bands, while the 24 $\mu$m light is powered by
younger more embedded stars which are more confined to the clumps.

For all wavelengths, the interacting galaxies have larger fractions in
clumps than the spirals.
On average, the median fractions for
the interacting galaxies are about a factor of 1.7 times larger than
those in the spirals.
This suggests that 
the star forming regions in the interacting galaxies are more
luminous, dustier, or younger on average than those in the spirals, so
are easier to detect at 8 $\mu$m.
The difference 
in the clump fraction
is likely not due to sensitivity differences,
as the 8 $\mu$m images of the spirals are on average more sensitive
to lower luminosity clumps
than those of the interacting galaxies (see Figures 7 and 8).

\subsection{Background/Foreground Sources }

Some of the objects 
classified as `disk' or `tidal' 
may actually be 
foreground stars, background quasars, or background galaxies.
There are three ways to distinguish such objects.
First, if an optical spectrum is available for the source, 
the redshift of the clump
can be used to confirm its association with the galaxy.
Unfortunately, for only a handful of the clumps 
are optical spectra
available, mainly galactic nuclei.

Second, a detection in the H$\alpha$ map
would confirm it is
at the same redshift as the galaxy, thus likely associated with the 
galaxy.  Unfortunately, however, H$\alpha$ images are not available
for all of the galaxies in our sample.  Even when an H$\alpha$ map
is available, it may not be sensitive enough to detect all of the 
star forming regions in the galaxy, particularly those that are
highly obscured and/or 
low luminosity. 

A third way to identify foreground or background interlopers
is via their position in Spitzer infrared color-color plots
\citep{smith05, lapham13}.     
Foreground stars have near-to-mid-infrared 
spectral energy distributions (SEDs)
that drop with increasing wavelength, 
star forming
regions have SEDs that drop from 3.6 $\mu$m to 4.5 $\mu$m and then 
increase at
longer wavelengths, and quasars have flat SEDs in this wavelength
range.
The application of this
method to our clump sample
is described in detail in the Appendix to this paper.

\section{Clump Star Formation Rates }

\subsection{Calculating SFRs}

For both the 2.5 kpc and 1.0 kpc
samples, we made six different estimates of
the SFR of each clump.
For clumps detected ($\ge$3$\sigma$) in both the FUV and 24 $\mu$m filters, we 
made a first estimate of the star formation rate, SFR$_{FUV+24}$, 
by 
correcting the FUV luminosity L$_{FUV}$ for extinction using
L$_{FUV}$(corrected) = L$_{FUV}$ + 6.0L$_{24}$ 
and 
then using the relation 
SFR$_{FUV+24}$ = 3.39 $\times$ 10$^{-44}$L$_{FUV}$(corr)
\citep{leroy08, liu11}.\footnote{Note that \citet{hao11}
derive a somewhat different relationship:
SFR$_{FUV+24}$ = 5.0 $\times$ 10$^{-44}$L$_{FUV}$(corr)
and L$_{FUV}$(corr) = L$_{FUV}$ + (3.89 $\pm$ 0.15)L$_{24}$ (see
also \citealp{kennicutt12}).
The \citet{hao11} conversion is intended for the global fluxes of galaxies,
while the \citet{leroy08} and \citet{liu11} relation is for spatially-resolved
studies (C. Hao 2015, private communication).
} 
In these formulae, the SFR is in units of M$_{\sun}$~yr$^{-1}$
and the luminosities are monochromatic luminosities ($\nu$L$_{\nu}$)
in erg~s$^{-1}$.

For systems with both NUV and 24 $\mu$m measurements,
we obtained a second estimate SFR$_{NUV+24}$ by 
correcting the NUV for extinction using
L$_{NUV}$(corr) = L$_{NUV}$ + 2.26L$_{24}$ 
and then using
SFR$_{NUV+24}$ = 6.76 $\times$ 10$^{-44}$L$_{NUV}$(corr)
\citep{hao11, kennicutt12}.

Third, for the clumps detected in H$\alpha$ and at 24 $\mu$m,
we obtain
another estimate of the SFR 
using the
equation SFR$_{H{\alpha}+24}$
= 5.5 $\times$ 10$^{-42}$[L$_{H\alpha}$ + 0.031~L$_{24}$].
This relationship was
found for H~II regions in nearby galaxies
assuming a Kroupa initial mass function
\citep{calzetti07, kennicutt09}.

Our fourth estimate of the SFR was obtained from the 24 $\mu$m
photometry alone, using 
SFR$_{24}$ = 2.0 $\times$ 10$^{-43}$L$_{24}$
\citep{rieke09}.  This is especially valuable for systems without
NUV, FUV, and H$\alpha$ images.

For systems with both NUV and 8 $\mu$m images, 
we make a fifth estimate of the SFR, SFR$_{NUV+8}$, by using 
L$_{NUV}$(corr) = L$_{NUV}$ + 1.24L$_{8}$.  We obtained
this relation by combining
the 8 $\mu$m and 24 $\mu$m relations of
\citet{kennicutt09}.
Finally, we obtained a sixth estimate of the SFR from the 
8.0 $\mu$m luminosity alone, 
using SFR$_{8}$ = 1.63 $\times$ 10$^{-43}$L$_{8}$,
after using the 3.6 $\mu$m luminosity L$_{3.6}$ to correct 
for contributions to
the 8 $\mu$m flux from starlight from the 3.6 $\mu$m flux
with
F$_8$(starlight) = 0.26F$_{3.6}$
\citep{wu05}.

The derived SFRs 
are a measure of the extinction-corrected H$\alpha$
luminosities of the clumps, and therefore the number of ionizing photons
in the region.
The SFR equations given above depend upon the assumed
dust attenuation law, which depends upon the dust distribution
relative to the UV-emitting stars and the ionized gas.
This issue is discussed further in Section 6.

These
estimates of SFR also depend upon the star formation history for the
region. 
These formulae
were derived assuming 
constant star formation rates over the last $\sim$10 $-$ 100 Myrs
(see \citealp{kennicutt12}), while the clumps have probably undergone recent
bursts of star formation.  
They are particularly uncertain for low luminosity clumps, when
stochastic effects can be important.  
In spite of these limitations and uncertainties,
however, these numbers aid comparison to other studies,
which frequently quote SFRs for individual knots of star formation
within galaxies
(e.g., \citealp{boquien07, boquien09b, boquien11, kennicutt07,
cao07, beirao09, pancoast10, smith14})
as well as galaxies as a whole (e.g., \citealp{kennicutt12}).

In general, the reliability of the SFR tracers given above depends upon
the band or bands used. 
In the following analysis,
we use
SFR$_{FUV+24}$ 
as our preferred estimate of the SFR 
when it is available
(36/46 interacting galaxies; 31/39
spirals).
If it is not available, 
in order of preference
we use
SFR$_{NUV+24}$ (4 interacting and 3 spiral galaxies), 
SFR$_{H{\alpha}+24}$ (4 interacting and 1 spiral), 
SFR$_{24}$ (1 interacting and 1 spiral),
SFR$_{NUV+8}$ (1 interacting and 3 spirals), 
and SFR$_{8}$ (no interacting and no spirals).
In general, 
the 24 $\mu$m flux is considered
a better tracer of star formation than 8 $\mu$m. The 24 $\mu$m light
from galaxies 
is dominated by emission 
from `very small interstellar dust grains'
heated mainly by UV light from massive young stars 
(e.g., \citealp{li01}).  
In contrast,
the 8 $\mu$m flux in star forming galaxies may be powered in 
part by older stars.
The 8 $\mu$m luminosity is not linearly proportional
to estimates of star formation from the near-infrared
hydrogen Pa$\alpha$ line, unlike
the 24 $\mu$m luminosity \citep{calzetti05, calzetti07}.
Further, since the 8 $\mu$m broadband Spitzer flux from galaxies
is dominated by spectral features
produced by 
polycyclic aromatic hydrocarbons (PAHs)
\citep{li01},
this flux can vary with metallicity, with low metallicity regions
being particularly deficient in the 8 $\mu$m band compared to 24 $\mu$m
\citep{boselli98, engelbracht05, rosenberg06, rosenberg08}.
Thus SFR tracers using 24 $\mu$m, if available, 
are preferred to those using 8 $\mu$m. 
Likewise, tracers using the FUV are preferred over those using the NUV, since
older stars contribute more to the NUV light (e.g., \citealp{kennicutt12}).
H$\alpha$ + 24 $\mu$m is generally considered a very reliable measure
of star formation (e.g., \citealp{kennicutt09}).
However, we rank SFR$_{H{\alpha}+24}$
lower in preference than estimates using the FUV or NUV with 24 $\mu$m
because our H$\alpha$ data are inhomogeneous, being obtained
from a variety of sources.
Using either UV or H$\alpha$ in addition to 24 $\mu$m is considered 
more accurate than 24 $\mu$m alone, because extincted
and unextincted young stars are both directly accounted for 
(e.g., \citealp{kennicutt09}).
Thus 24 $\mu$m alone is only used when UV or H$\alpha$ measurements 
are not available.  Only when no alternative is available
do we use the 8 $\mu$m alone as our indicator of the SFR.

In Figures 5 and 6, we plot various ratios
of the different measures of star formation against SFR.
These plots indicate
reasonably good consistency between the different methods
of determining SFR, but with some scatter.
Some of the scatter in Figures 5 and 6 is likely due to how
the different SFR tracers depend upon properties of
the region such as age, star formation history,
dust composition, extinction, geometry, and initial
mass function. For example, FUV is skewed towards younger stars
than the NUV, and the UV is more sensitive to the assumed dust properties 
than H$\alpha$.
Note that there is a
slight anti-correlation between SFR$_{NUV+24}$/SFR$_{FUV+24}$ and 
SFR.  This may be due to clump-to-clump age variations, 
with younger ages giving 
both 
lower SFR$_{NUV+24}$/SFR$_{FUV+24}$ and higher SFRs.
Alternatively, the attenuation correction may vary
with SFR. 
For low luminosity regions, stochastic variations from region
to region due to random sampling of the initial mass function may
also play a role, however, this is not a factor for the high
SFR regions.

Our ranking of the various tracers of SFR is supported
by Figures 5 and 6.
Moderate
scatter 
is seen for the ratio SFR$_{NUV+24}$/SFR$_{FUV+24}$, showing that these
two methods give reasonably consistent SFRs.
In contrast, large scatter is seen for 
SFR$_{NUV+8}$/SFR$_{FUV+24}$
and 
SFR$_{8}$/SFR$_{FUV+24}$, supporting the assertion that 8 $\mu$m, with
or without UV, 
is a less reliable indicator of SFR.
Ratios involving H$\alpha$ also show moderate scatter.
The SFR$_{24}$/SFR$_{NUV+24}$ ratios are skewed towards values less than
one, likely because unobscured young stars are not being included in
the census in SFR$_{24}$.

We note that the above SFR relations 
depend upon metallicity, especially those involving
the 8 $\mu$m flux, particularly at oxygen abundances
less than log(O/H) + 12 = 8.4
\citep{calzetti07}.
No abundance determinations are available for most of the star forming
regions in our sample.  In most cases, however, they are likely to be
higher than this limit.
For example, the tidal features
in Arp 72, Arp 105, and Arp 245 have 
log(O/H) + 12 = 8.7, 8.6, and 8.65, respectively
\citep{smith10, duc94, duc00}.

\subsection{Histograms of SFRs}

Figure 7 displays histograms of the SFRs of the clumps obtained with
a 2.5 kpc radius aperture.
These histograms are proxies for the 
extinction-corrected 
H$\alpha$ luminosity functions of the regions.
The top panel in Figure 7 shows the 
disks and tidal clumps in the interacting galaxies, while
the disk clumps of the spiral galaxies are given in the second panel.
The hatched histogram in the top panel indicates the tidal clumps.
Similar histograms for the 1.0 kpc radius clumps are provided
in Figure 8.

Note that Figures 7 and 8 do not include the `nuclear clumps', just the extranuclear
`disk' and `tidal' regions.   The fluxes
from the nuclear regions may be contaminated
by active galactic nuclei.  In addition,
the stellar populations in the nuclear and immediate 
circumnuclear regions may differ
from those in star forming regions further out in the disk.  
For example, circumnuclear rings
may have long
star formation histories in comparison to regions further out in the disk 
(e.g., \citealp{kennicutt89c, sarzi07, dors08}).  
In our sample galaxies, such circumnuclear rings (for example, in NGC 1097 and NGC 4321)
are contained within the `nuclear clumps', thus do not contribute to Figures 7 and 8. 

Along the top axes of Figures 7 and 8, 
we converted the SFRs into SFR per area
by dividing by the area per aperture
(19.6 kpc$^2$ for the 2.5 kpc
aperture radii regions and 3.14 kpc$^{2}$ for the 1.0 kpc aperture
radii clumps).
In these plots, the red histograms
mark
the clumps that are detected in H$\alpha$, and thus are confirmed to be
associated with the galaxies.
In the blue histograms,
we identify
the clumps
that are {\it either} detected in H$\alpha$ 
or do {\it not} lie in the
stellar or quasar areas in the Spitzer color plots, or both (see Appendix).
In other words, clumps not included in the blue histogram are 
likely {\it not} associated with the galaxies.
Clumps in the blue histogram but not in the red histogram may be part of
the galaxies, but that is uncertain.

In all cases, the histograms of SFRs are centrally peaked.
The drop off in the number of clumps with lower luminosity
may be due in part to incompleteness.   We did two calculations
to estimate the completeness limit of the sample.
First, for the smoothed 8 $\mu$m images, we 
calculated the theoretical sensitivity to point sources, based on the
10$\sigma$ clump selection criteria used by {\it daofind}.
Histograms of these limits, after conversion to SFR, 
are given in the lower two panels in Figures 7
and 8, 
for the 2.5 kpc and 1.0 kpc samples,
respectively.
The y-axes on these lower histograms are the logarithm of the 
number of galaxies with each sensitivity.
Figures 7 and 8 
show that the turn-over in the luminosity function
is significantly higher than the theoretical 10$\sigma$
sensitivity limit for the
images.

However, crowding of clumps may lead
to blending and therefore incompleteness.  
To test for this, we used the IRAF {\it mkobjects} routine to randomly
add
artificial clumps to the disks of the galaxies in 
our smoothed 8 $\mu$m images.
We then determined how many of these sources were recovered
by {\it daofind}.
For the 1.0 kpc radius clump sample, we 
find the sample is 
74\% complete at 
a SFR = 0.025 M$_{\sun}$~yr$^{-1}$, 
80\% complete at 
a SFR = 0.04 M$_{\sun}$~yr$^{-1}$, 
90\% complete at 
a SFR = 0.1 M$_{\sun}$~yr$^{-1}$, 
and 
95\% complete at 
a SFR = 0.25 M$_{\sun}$~yr$^{-1}$.
For the 2.5 kpc sample, we find an estimated 
53\% completeness at 0.16
M$_{\sun}$~yr$^{-1}$,
60\% completeness at 
0.25 M$_{\sun}$~yr$^{-1}$,
70\% at 
0.4 M$_{\sun}$~yr$^{-1}$, and 
85\% 
at SFR = 1 M$_{\sun}$~yr$^{-1}$.

For both samples of clumps, we fit the high SFR end of the 
log N - log SFR distribution
to a straight line, where N is the number of clumps in a luminosity
bin and the width of the bin scales with log(SFR) as in Figures 7 and 8.
For the 2.5 kpc samples,
we only fit the distribution above log(SFR) $>$ $-$0.8 
(SFR $>$ 0.16 M$_{\sun}$~yr$^{-1}$), 
above the point where the distribution function turns over.
For the 1.0 kpc sample, we used a cut-off of log(SFR) $>$ $-$1.4
(SFR $>$ 0.04 M$_{\sun}$~yr$^{-1}$).
For these fits, we only included the data within the blue histograms,
thus we are eliminating the most likely foreground/background
interlopers from the sample.  Adding the excluded sources back
in the sample doesn't change the results significantly, since
very few of the sources above our cutoffs lie in the stellar or
quasar area of the Spitzer color-color plot.

The best-fit slopes of these lines are provided in Figures 7 and 8.  
For the 2.5 kpc sample, we find a slope of 
$-$0.79 $\pm$ 0.14 for the 
interacting galaxies and $-$1.01 $\pm$ 0.15 for the spiral galaxies.
For the 1.0 kpc sample, the slope is
$-$0.92 $\pm$ 0.06 for the 
interacting galaxies and $-$1.54 $\pm$ 0.13 for the spiral galaxies.
For both samples of clumps, the distribution of SFR
is flatter for the interacting clumps
than for the spiral clumps, 
that is, there are proportionally more
high SFR clumps in the interacting galaxies than in the spirals (Figures
7 and 8).
For the 2.5 kpc sample, a Kolmogorov-Smirnov test on the
distribution of clumps above 
log(SFR) $>$ $-$0.8
(SFR $>$ 0.16 M$_{\sun}$~yr$^{-1}$) gives a probability
of 2\% that the two sets of clumps come from the same parent sample.
For the 1.0 kpc sample, 
using a Kolmogorov-Smirnov test on the distribution above
log(SFR) $>$ $-$1.4
(SFR $>$ 0.04 M$_{\sun}$~yr$^{-1}$) gives a probability
of 1.6\% that the interacting and spiral
clumps come from the same parent sample.

To test whether systematic differences between the various SFR indicators
(Figures 5 and 6) affects these conclusions, we  
repeated this analysis using a different
order of preference for the 
SFR indicators.
In this second analysis, we 
used 
SFR$_{H{\alpha}+24}$ as the top choice if available, followed by
SFR$_{NUV+24}$, 
SFR$_{FUV+24}$, 
SFR$_{24}$, 
SFR$_{NUV+8}$, 
and SFR$_{8}$ in order.
We then fit the high SFR end of the SFR distribution to power laws
as in Figures 7 and 8.
For the 2.5 kpc sample, we find a slope of 
$-$0.64 $\pm$ 0.37 for the 
interacting galaxies and $-$0.82 $\pm$ 0.24 for the spiral galaxies.
For the 1.0 kpc sample, the slope is
$-$0.90 $\pm$ 0.09 for the 
interacting galaxies and $-$1.45 $\pm$ 0.18 for the spiral galaxies.
These are consistent with the results found for the original 
priority ordering scheme though with larger uncertainties.

The results shown in Figures 7 and 8 
imply that there are more young stars, on average, per region
in the interacting galaxies (i.e., a higher density
of stars).  This means that either the efficiency of star formation
per region is higher in interacting
galaxies, or there is more interstellar gas per clump,
or both.
This issue is discussed further in Section 6.3.

An alternative explanation for 
the higher SFRs for the clumps in the
interacting galaxies is that clumps are `blended' together more
frequently in the interacting galaxies, which are more distant on average.
However, we are extracting the photometry over the same physical
scale in all of the galaxies, so
the relative amount of blending should be similar in the two
samples, unless star forming regions
are closer together on average in the interacting
galaxies or more frequently aligned along our line of sight in
the interacting galaxies due to
the tidal interaction.
One argument against blending alone being responsible 
for the difference between the
two samples comes from high resolution Hubble Telescope images
of the highest SFR regions.  As discussed in 
\citet{smith14} and summarized in Section 5.3 of the current paper,
at high
resolution the
most luminous clumps in our sample 
are seen to contain extremely luminous star clusters at their core.

For a 2.5 kpc radius aperture clumps,
the slope of the
tidal clumps in the interacting galaxies
(the hatched histogram in Figure 7) show a flatter
distribution than the disk clumps in the interacting galaxies, 
indicating that
the highest SFR regions are preferentially tidal. 
However, this is uncertain because the 
classification of `disk' vs.\ `tail'
is quite ambiguous.
In contrast, most of the highest SFR regions in the 1.0 kpc sample are
classified as disk sources (Figure 8).
As discussed below, the distance-limited
1.0 kpc sample lacks some of the most extreme
regions (i.e., the hinge regions in Arp 240 and Arp 256).

\subsection{The Highest SFR Clumps}

A comparison of the top and bottom panels in Figures 7 and 8 shows
that the most luminous clumps are found in the interacting galaxies.
On a 2.5 kpc radius scale (Figure 7), 
no clumps are found in the spiral sample
with SFR $\ge$ 2 M$_{\sun}$~yr$^{-1}$, 
while five such clumps are found in the interacting
sample.  These include three hinge clumps in Arp 240 
(3.3, 6.2, and 9.0 M$_{\sun}$~yr$^{-1}$),
the `overlap region' between the two disks of the Antennae galaxies
Arp 244 (4.8  
M$_{\sun}$~yr$^{-1}$), 
and the hinge clump
in Arp 256 (2.0 
M$_{\sun}$~yr$^{-1}$).   A knot of star formation near the
base of the short western tail of NGC 2207 
has a slightly lower SFR
of 1.8
M$_{\sun}$~yr$^{-1}$.
These regions and their host galaxies
are discussed in detail in \citet{smith14}.

Hubble Telescope images
reveal extremely luminous
star clusters
(M$_{\rm I}$ between $-$12.2 and $-$16.5) at the
centers 
of the Arp 256, Arp 240, and NGC 2207 clumps
\citep{smith14}.  
These clusters are resolved with
HST with estimated sizes of $\sim$70 pc, much larger 
than most 
luminous star clusters (e.g., \citealp{larsen04}).
If individual star clusters, their luminosities would 
place them near the top of the luminosity function
for extragalactic star clusters (e.g., \citealp{gieles10}).
These sources may actually be 
tightly
packed groups of star clusters, rather than individual clusters.
HST images of the more nearby Antennae galaxies shows tightly
grouped complexes of star clusters in the overlap region
\citep{whitmore2010}.
If smoothed to the effective resolution of Arp 240 and Arp 256,
these clusters would be blended together
to produce a source similar in size and luminosity to the
sources seen in those galaxies \citep{smith14}.
Alternatively, these sources may be single very large clusters.
Numerical simulations of star cluster formation and evolution
in galaxy interactions indicate that large clusters can
be produced by either cluster mergers, mergers of gas clouds,
or expansion due to a passage through the inner disk of 
the galaxy \citep{fellhauer05, renaud14}.

In an early study, \citet{kennicutt88b}
found that some of the largest H~II regions
in nearby galaxies host unusually luminous star clusters.
Later studies have indicated that the global star 
cluster formation efficiency in galaxies
(the fraction of star formation occurring in dense bound star clusters) 
increases as the star formation rate of the galaxy increases
\citep{goddard10, adamo11, adamo15}.
Locally within galaxies this trend is also seen, along with a trend
towards higher cluster formation efficiency in regions with higher
gas surface density \citep{ryon14, adamo15}.
Theoretical models \citep{elmegreen08, kruijssen12} predict such a trend,
as regions with higher gas density have shorter free-fall times
and higher star formation efficiencies, thus are less affected
by gas expulsion and so are more likely to form dense bound clusters.
Our observation of very luminous clusters or dense clusters of clusters
in the clumps with the highest SFRs may be consistent with this trend.

Another intriguing difference between the highest SFR
knots and those in normal spiral galaxies comes
from Chandra X-ray images.   In Chandra maps, the highest SFR
regions tend to have strong diffuse X-ray 
emission with a soft spectrum, indicative of hot gas
\citep{smith14}.
The diffuse L$_{\rm X}$/SFR ratios for the highest SFR clumps
are higher than those for regions in normal spirals,
suggesting higher density gas
on average
\citep{smith14}.  This is consistent with a picture in which
large quantities of gas are being driven into these regions.

In the spiral sample at the 2.5 kpc radius scale, only two clumps
have SFRs greater than 1 M$_{\sun}$~yr$^{-1}$.
These both 
lie
in the outer spiral
arms of NGC 3646, and have inferred SFRs of 
1.2 and 1.5 M$_{\sun}$~yr$^{-1}$
(see Figure 9).
Although it is in our control sample of `normal' spiral galaxies,
NGC 3646 does have a companion, NGC 3649, which has a blue
luminosity of 1/24$^{th}$ that of NGC 3646 (from NED).
This is just below the cut-off used by 
\citet{smith07, smith10}
to eliminate interacting galaxies from the control sample. 
Thus NGC 3646 is interacting, though not as strongly as the systems
in the SB\&T sample.
Interestingly, 
the most luminous clumps in NGC 3646 
lie near where spiral arms appear to cross,
with a morphology suggestive of intersecting caustics.
Thus at least some of the star formation that is occuring
in our control sample was likely also triggered by interactions.
Numerical simulations of flyby encounters between galaxies
shows that low mass companions (0.01 $-$ 0.1 $\times$ the mass
of the primary) can induce long-lived tidal waves in galaxies
\citep{struck11}.

NGC 3646 has been identified as being one of the largest and 
most luminous spirals
in the local Universe \citep{vandenbergh60, romanishin83}, with
a D$_{25}$ diameter 
\citep{devaucouleurs91}
of 74 kpc at a distance of 65 Mpc.
Thus even if it was not interacting
it would be a likely host for luminous star forming regions.
For spiral galaxies of a given Hubble type, 
the luminosities of the brightest H~II regions scale
with the luminosity of the galaxy as a whole \citep{kennicutt88}.
NGC 3646 has the 3rd largest
3.6 $\mu$m luminosity of the spiral galaxies in our sample (see Figure 2).

As discussed in Section 4, due to
resolution issues the 1.0 kpc sample of clumps is limited
to galaxies within 67 Mpc.   
That means that two of the galaxies with the most luminous 
clumps in the 2.5 kpc sample, Arp 240 and Arp 256, 
are not included in the 1.0 kpc clump sample.
However, 
Figure 8 shows that 
even without these more distant systems
there is a difference between
interacting and normal galaxies.
On a 1.0 kpc radius scale,
interacting galaxies host proportionally more
clumps with log(SFR) $>$ $-$0.6 (SFR $>$ 0.25 M$_{\sun}$~yr$^{-1}$) compared to
the spiral galaxies (Figure 8).

In the 1.0 kpc sample, 
the clump with the highest SFR
is the overlap
region in the Antennae galaxies Arp 244.
The two NGC 3646 clumps are the most luminous in the spiral sample.
The derived luminosities of these clumps are similar with the two
aperture sizes, indicating that the star formation is compact
relative to the 2.5 kpc radius aperture.

\subsection{Comparison to Earlier Studies} 

In a now classic paper, \citet{kennicutt89} derived H$\alpha$ luminosity
functions of HII regions in a sample of nearby spiral and irregular
galaxies on scales of 100 $-$ 200 pc.  
They found a trend with Hubble type,
with flatter luminosity functions for the Sm $-$ Im galaxies than for 
Sab $-$ Sb galaxies on average, with Sbc and Sc in between.
Our best-fit slope for the 1.0 kpc sample of clumps in 
our interacting galaxies 
is consistent with their results for their Sm $-$ Im galaxies,
while our best-fit value for our spiral clumps is 
in the range for their spirals
(note that they fit dN/dL = AL$^\alpha$, while we fit to 
dN/d(log(L)), thus our slope = $\alpha$ + 1).  
Our 2.5 kpc sample for the interacting galaxies gives a slope between
that of the Sm $-$ Im galaxies and the spirals, while the 
slope for the 
2.5 kpc spiral sample 
is similar to that of early-type spirals.

In a recent study, \citet{zaragoza15}
use H$\alpha$ Fabry-Perot data 
to derive H$\alpha$ luminosities and velocity dispersions for 
H~II regions in 
12 galaxies in 
nine interacting systems, and compared with 28 isolated galaxies.
Similar to our results, 
they found a flatter H~II region luminosity function for the 
interacting systems than for the normal galaxies.   
Their study involved smaller spatial scales than ours
(the diameters of their regions ranged from 15 $-$ 400 pc)
and their H$\alpha$ fluxes were uncorrected for attenuation,
but the slopes they found for the high luminosity end of their
luminosity functions are similar to ours.

\section{Dust Attenuation }

\subsection{Dust Attenuation vs.\ SFR}

In Figures 10, 11, and 12, we compare the SFRs of the clumps against
three standard indicators of dust extinction: 
the L$_{H\alpha}$/L$_{24}$
ratio, the NUV $-$ [24] color, and the FUV $-$ [24] color, respectively.
In these plots, the 2.5 kpc radius
clumps are shown in the top panel, and the 1.0
kpc radius clumps in the bottom.
In all cases, we see a trend, such that the higher SFR clumps have more
dust absorption.  

The best linear fits to the datapoints are shown in Figures 10, 11, and
12 as solid blue lines.  
Inspection of these plots shows that the trends are not 
exactly linear, but are flatter at the low SFR end, steepening 
at higher SFR.
Note that there is quite a bit of scatter about the best-fit
lines, particularly
for the 1.0 kpc scale clumps, for which no correlation is visible if clumps
with log SFR $>$ $-$1.4 (SFR $>$ 0.04 M$_{\sun}$~yr$^{-1}$) are excluded.
On average, the clumps in the spiral galaxies have less
dust absorption than those in the interacting galaxies,
but follow the same general trend.

The observed
correlations between SFR and implied dust attenuation 
may occur because regions with higher gas column densities
have higher dust column densities, 
and higher gas column densities produce higher SFRs.
The latter relationship has been quantified
in the so-called Schmidt-Kennicutt Law 
\citep{schmidt59, kennicutt89b,
kennicutt98},
in which the SFR surface density
$\Sigma$$_{SFR}$ 
is related to the surface density of the 
interstellar gas $\Sigma$$_{gas}$ via a power law.
This relation is discussed further in Section 6.3.

A correlation between the apparent UV attenuation 
of star forming regions 
and their luminosity
was found before by 
\citet{boquien09}.  
They extracted GALEX FUV and Spitzer 24 $\mu$m fluxes for 
regions within eight nearby spiral galaxies, using a spatial scale of
$\sim$300 pc.
They found that L$_{24}$ increases at a faster rate than 
L$_{\rm FUV}$, consistent with our Figure 12.

\subsection{Dust Attenuation Laws}

In Figure 13, we compare 
our three indicators of dust attenuation against each other.
Again, the datapoints are color-coded as 
open black squares for the disk clumps in the interacting galaxies,
magenta open diamonds for the tidal clumps,
and 
small blue filled triangles for clumps in the disks of the
spirals.
Regions with SFRs greater than 0.4 M$_{\sun}$~yr$^{-1}$
have a green circle around them.
In general, the three tracers of attenuation correlate with each other,
though with significant scatter. 
The observed trends in these plots
provide us with a measure of the average attenuation law 
in these regions, while 
the observed scatter around the average trends is likely due to 
variations in the geometry, age, and dust properties of the clumps.
In Figure 13, the solid magenta lines show the best 
linear fits to
the data for all of the clumps, while
the green dotted lines are the best-fit lines for the clumps
with SFR $>$ 0.4 M$_{\sun}$~yr$^{-1}$.

The high SFR clumps trace out a steeper path in Figure 13
than 
the full sample of clumps.
This implies that for the high SFR clumps, for a given change
in the observed H$\alpha$, the UV changes more, compared to
lower luminosity clumps. 
This may be due to a steeper UV-to-H$\alpha$
dust attenuation curve in high SFR clumps than in lower luminosity clumps.
Alternatively, it may be caused by age differences between the clumps.
These possibilities are discussed further below.

Our dataset can be used to 
estimate the relative 
dust attenuation in 
the H$\alpha$ and UV filters.
A standard technique for
deriving dust attenuation laws 
is the 
`SFR matching' method, as outlined by \citet{hao11} (also see
\citealp{buat02},
\citealp{treyer07} and \citealp{boquien15}).
This starts with the generic formula
for the SFR based on the UV, NUV, or H$\alpha$
luminosities as used in Section 5.1:

{\center SFR = C$_{\lambda}$L$_{\lambda}$(corrected) }

\noindent where 
L$_{\lambda}$(corrected) is the luminosity in the band
after correction for attenuation, and 
C$_{\lambda}$ is the conversion factor to SFR for that 
band.  
L$_{\lambda}$(corr) is then equal to 
L$_{\lambda}$(observed)10$^{0.4A_{\lambda}}$,
where A$_{\lambda}$ is the effective attenuation in 
that band in magnitudes.
We then assume that the H$\alpha$ attenuation
obtained using the formula for SFR$_{{H\alpha}+24}$
given in Section 5.1 
is reliable:  
A$_{H\alpha}$ = 2.5log(1 + 0.031L$_{24}$/L$_{H\alpha}$)
(from \citealp{calzetti07}
and \citealp{kennicutt09}).\footnote{We note that \citet{zhu08}
find A$_{H\alpha}$ = 2.5log(1 + 0.020L$_{24}$/L$_{H\alpha}$)
As discussed in \citet{zhu08} and \citet{kennicutt09}, the \citet{zhu08}
formula is for galaxies as a whole,
while the \citet{calzetti07} relation is for spatially-resolved observations.
}

We then set the SFRs derived from other filters
equal to that from the corrected H$\alpha$, i.e.:

{\center SFR$_{\lambda}$ = SFR$_{H{\alpha}}$(corr),}

where SFR$_{H{\alpha}}$(corr) =
C$_{H\alpha}$L$_{H\alpha}$(obs)10$^{0.4A_{H\alpha}}$.

Rearranging and plugging in $\lambda$ = FUV gives:

log(L$_{H\alpha}$/L$_{\rm FUV}$)(observed) = 
0.4(A$_{\rm FUV}$/A$_{H\alpha}$ $-$ 1))A$_{H\alpha}$ - 
log(C$_{H\alpha}$/C$_{\rm FUV}$)

According to the above formula,
if 
A$_{\rm FUV}$/A$_{H\alpha}$ does not vary from clump to clump
and the SFR matching technique is valid for these regions,
the slope of the best-fit line in a plot of 
log(L$_{H\alpha}$/L$_{\rm FUV}$)(observed) vs.\ A$_{H\alpha}$
should equal 
0.4(A$_{\rm FUV}$/A$_{H\alpha}$ $-$ 1)).

Note that the SFR matching method
assumes that the SFR is approximately constant
over a timescale of $\sim$100 Myrs.  H$\alpha$ is most sensitive
to star formation on a timescale of about 10 Myrs, since only the
most luminous stars produce large quantities of hydrogen-ionizing UV flux.
In contrast, the FUV is dominated by light from somewhat lower mass
stars, so measures star formation over a timescale of $\sim$100 Myrs,
and NUV even longer timescales, to $\sim$200 Myrs \citep{kennicutt12}.
Thus variations in the SFR over short timescales, such as recent
bursts, can significantly
affect results from the SFR matching method.
This point is discussed further below.

In Figure 14, we plot the observed
log(L$_{H\alpha}$/L$_{\rm FUV}$) vs.\ A$_{H\alpha}$
(bottom panels).  Figure 14 also provides plots of 
the observed
log(L$_{H\alpha}$/L$_{\rm NUV}$) vs.\ A$_{H\alpha}$
(top panels).
The 1.0 kpc radius clumps are displayed in the left panels of Figure 14,
while the 2.5 kpc clumps are in the right panels.
The 
open black squares are the disk clumps in the interacting galaxies,
the 
magenta open diamonds are the tidal clumps,
and the 
small blue filled triangles are clumps in the disks of the
spirals.
Regions with SFRs greater than 0.4 M$_{\sun}$~yr$^{-1}$
have a green circle around them.

On Figure 14, we overlay the 
standard 
\citet{calzetti01} starburst attenuation law (red dashed curve),
assuming that the FUV and NUV attenuations also obey
the relations relative to 24 $\mu$m 
implied by the formulae for SFR$_{FUV+24}$
and SFR$_{NUV+24}$ given above, i.e.,  
A$_{FUV}$ 
= 2.5log(1 + 6.0L$_{24}$/L$_{FUV}$)
\citep{leroy08, liu11} and 
A$_{NUV}$ 
= 2.5log(1 + 2.26L$_{24}$/L$_{NUV}$)
\citep{hao11}.
The 
\citet{calzetti01}
starburst law
gives 
A$_{\rm FUV}$ = 1.82 A$_{H\alpha}$ and 
A$_{\rm NUV}$ = 1.52 A$_{H\alpha}$
(see \citealp{hao11} and \citealp{boquien15}).
\citet{hao11} 
found a similar 
A$_{FUV}$/A$_{H\alpha}$ of 2.06 $\pm$ 0.28 
from the global fluxes of nearby star forming galaxies.

There is a large amount of scatter in the data plotted in Figure 14,
without strong overall trends.
However, there are differences between the high SFR regions
and the rest of the clumps.
In these plots,
the high SFR clumps generally follow paths that trend upwards,
and the data points tend to lie near or along the Calzetti relation.
In contrast, the lower SFR regions show a slight trend downwards.
A downward slope in these plots
means that as the overall attenuation increases,
the attenuation in H$\alpha$ increases faster than 
that in the NUV.  

We thus conclude that the Calzetti attenuation law is a reasonable match
to the data for the high SFR clumps.
In contrast, the downward trends for the lower luminosity regions 
imply smaller 
A$_{FUV}$/A$_{H\alpha}$ and 
A$_{NUV}$/A$_{H\alpha}$ ratios than the Calzetti law, if the SFR
matching method is appropriate for those clumps.

As noted above, the SFR matching method requires an approximately constant
SFR over timescales 
of 10 $-$ 100 Myrs.
For the highest SFR clumps in our sample, 
we have conducted detailed stellar population synthesis studies,
and found that the broadband
optical/UV/IR colors and the H$\alpha$ data are inconsistent with
a single instantaneous burst \citep{smith14}.  
Instead, population synthesis 
suggests 
that either these regions have undergone two or more bursts, or the 
star formation in these regions has been prolonged over an extended 
time period.  
The presence of large quantities of diffuse X-ray-emitting gas
in these extreme regions \citep{smith14} supports this scenario.
Thus, at least for the highest luminosity regions
in our sample, an assumption of a constant SFR over $\sim$100 Myrs
may be a reasonable first approximation in estimating the dust
attenuation laws.  This supports the
idea that the Calzetti law is appropriate for these high luminosity
regions.

The Calzetti law was derived for global fluxes
from starburst galaxies, 
and 
can be explained by a clumpy foreground
dust screen \citep{calzetti96, calzetti97,
charlot00,
fishera03}.
Such a geometry gives
a flatter attenuation curve with wavelength 
(more `gray')
and a larger ratio
of total to selective extinction
than a uniform dust screen \citep{natta84, witt00}.

We conclude that 
our highest luminosity clumps are 
starburst-like in their dust attenuation.  This suggests that they
are also starburst-like 
in their dust distributions, meaning they are surrounded by
clumpy dusty gas shells.   
Perhaps in these regions star formation is occurring inside-out,
with a dense shell of gas and dust around a compact group of
highly obscured hot young stars.
Perhaps large quantities of
gas and dust are inflowing into these regions, contributing to dust
attenuation, but not yet involved in the star formation itself.

In starbursts, the ionized gas 
(i.e., the H$\alpha$)
has been found to be proportionally
more attenuated than the stars (i.e., the FUV and NUV), likely
because the ionized gas is more deeply embedded within dusty interstellar
clouds
than the overall stellar population
\citep{calzetti94, calzetti01}.  The differential attenuation
of the stars relative to the ionized gas in starburst
galaxies has been found to be about a factor of 0.44
\citep{calzetti94, calzetti01}.  

If SFR matching holds in these regions, our results imply higher
A$_{\rm FUV}$/A$_{H\alpha}$ ratios for higher SFRs.
A similar conclusion was reached by \citet{boquien15} from a 
high resolution study of the nearby spiral galaxy M33.
For the highest SFR regions ($\Sigma$$_{SFR}$
$\ge$ 0.1 M$_{\sun}$~yr$^{-1}$kpc$^{-2}$) at
the highest spatial scale (33 pc), 
\citet{boquien15} 
find 
A$_{\rm FUV}$/A$_{H\alpha}$ = 3.94, considerably higher than
the Calzetti value.
\citet{boquien15} interpret this increase
as being due to 
less differential attenuation
of the stars and gas on that spatial scale within intense star forming
regions.   
On that scale, 
the stars emitting the FUV light
are more likely to be associated with the star formation
and are therefore more mixed with the ionized gas and 
so 
attenuated by a similar amount of dust.
In general, 
on small spatial scales within M33, 
\citet{boquien15} find a trend in that regions with higher SFR have 
larger 
A$_{FUV}$/A$_{H\alpha}$, suggesting less differential extinction
between the stars and the gas.

In contrast to the high SFR regions, the lower luminosity
clumps in our sample have increasingly smaller 
L$_{H\alpha}$/L$_{\rm UV}$ ratios for increasing H$\alpha$ 
attenuation (Figure 14), implying that as the dust attenuation increases
the attenuation in the H$\alpha$ increases at a faster rate
than that in the UV.   
This trend is consistent with what is found on a global scale
in star forming galaxies by \citet{wild11}; as the stellar-mass-normalized
star formation rate drops, the differential attenuation between the
stars and the gas increases.
In other words, at lower SFRs, the fraction of the UV-emitting
stars still embedded in the interstellar cloud decreases.

In general, if the UV-emitting stars are more mixed 
with the ionized gas, with the attenuating dust outside of the bulk
of these stars, 
the effective A$_{FUV}$/A$_{H\alpha}$ 
and A$_{NUV}$/A$_{H\alpha}$ 
ratios increase,
while systems with more differential attenuation between the stars
and the ionized gas have 
lower ratios. 
Thus one possible explanation for our results for the
low luminosity regions is that there is more
differential extinction between the UV-emitting stars and the 
ionized gas in our clumps than in the Calzetti
starburst
relationship.
In general, a clump in which the stars are more
extended than a central concentration of interstellar gas
will produce a `flatter' attenuation curve (i.e., a lower 
A$_{FUV}$/A$_{H\alpha}$ ratio) \citep{witt00}.

Alternatively, the lower apparent
A$_{FUV}$/A$_{H\alpha}$ ratios
for the lower SFR clumps
may be due to non-constant SFRs in the clumps.
As noted above, by using the SFR matching technique we are implicitly
assuming that the SFR is constant over the timescale that these
tracers are sensitive to star formation (100 Myrs).   This is a reasonable
assumption for a galaxy as a whole, but less valid for an individual
knot of star formation.   If instead there was an
instantaneous burst of star formation in the knot, then the H$\alpha$ will
fade first (with a timescale of about 10 Myrs), then the FUV, then
the NUV.   
Instead of stronger 
attenuation
in the H$\alpha$ compared to the FUV, the low
L$_{H\alpha}$/L$_{\rm FUV}$ ratios may be due to 
an aged stellar population in which 
the intrinsic
L$_{H\alpha}$/L$_{\rm FUV}$ has dropped.
In a detailed study of the star forming regions in M51,
\citet{calzetti05} reached similiar conclusions: either 
the dust attenuation
law in the lower luminosity regions is different from that of starbursts,
or older ages are affecting the UV colors and UV/IR ratios.

Another factor that can strongly affect the 
A$_{FUV}$/A$_{H\alpha}$ ratios 
is the inclination of the disk.
Based on detailed population synthesis modeling and radiative transfer
calculations, \citet{wild11} found that disk galaxies more inclined
to our line of sight have larger differential attenuation between
the stars and the ionized gas; in the most extreme cases
the attenuation of the ionized gas can be more than four times 
that of the stars.
In interacting galaxies, tidal distortions 
may exacerbate the problem, 
placing unrelated tidal
features in front of other knots of star formation.

Other factors that may affect the 
observed L$_{H\alpha}$/L$_{\rm FUV}$ ratios are the metallicities,
the amount and density of the gas,
and the amount of H$\alpha$ leakage out of the region.
These possibilities could be tested with follow-up optical 
and UV spectroscopy and detailed population synthesis.

\subsection{The Schmidt-Kennicutt Law}

As noted earlier, the observed
correlations between SFR and dust attenuation (Figures 10 $-$ 12)
may exist because regions with higher gas column densities
have higher dust column densities, and the SFR is related
to the gas column density via the Schmidt-Kennicutt Law.
Globally, the disk-averaged relationship is
$\Sigma$$_{SFR}$ 
$\propto$ $\Sigma$$_{gas}$$^{1.4 \pm 0.15}$
\citep{kennicutt89b}.
Spatially-resolved studies find
similar relations within galaxies, with the power law index
varying somewhat depending upon the galaxy and spatial scale studied
\citep{kennicutt07, liu11}.

Unfortunately, 
high spatial
resolution maps of tracers of the interstellar
molecular gas (such as 2.6 mm CO emission) are not available
for most of the galaxies in our sample.  This means that 
we cannot study the Schmidt-Kennicutt Law directly in our
systems.   
However, we can investigate it indirectly, via its effect on the
dust attenuation.  
In external galaxies, the relationship between
the dust attenuation and the gas column density in star forming
regions 
is complicated, 
depending upon 
the geometry of the region, the metallicity,
the properties of the dust, and the gas column density itself.

As an empirical method of determining the relationship between
the dust attenuation and the hydrogen column density within galaxies,
we use the table of 
spatially-resolved measurements 
of the H$\alpha$ flux, the 24 $\mu$m flux density, and 
total hydrogen (HI + H$_2$) gas column density
N$_{\rm H}$ in 500 pc diameter regions within M51 
provided by \citet{kennicutt07}.
In Figure 15, 
we display their data in 
a plot of 
log L$_{H\alpha}$/L$_{24}$ vs.\ N$_{\rm H}$. 
A correlation is seen.  A linear fit (cyan line) gives
log L$_{H\alpha}$/L$_{24}$ = ($-$0.34 $\pm$ 0.05)(log N$_{\rm H}$)
+ (5.5 $\pm$ 1.0). 

We combine this equation with the spatially-resolved
Schmidt-Kennicutt law 
$\Sigma$$_{SFR}$ 
(in M$_{\sun}$~yr$^{-1}$kpc$^{-2}$)
= 4.8 $\times$ 10$^{-5}$ $\Sigma$$_{gas}$$^{1.56}$
derived by
\citet{kennicutt07} for 500 pc diameter regions within M51.
The resultant relationship is overplotted on our 
L$_{H\alpha}$/L$_{24}$
vs.\
SFR plots in Figure 16 as cyan dashed lines.
We see good agreement between these curves and our best-fit lines,
implying that the Schmidt-Kennicutt law holds in our regions.

To investigate this issue further, we utilize the 
relationship between FUV extinction and hydrogen column density
derived by \citet{boquien13} from a multi-wavelength
spatially-resolved study 
of a sample of four very nearby spiral galaxies
at a resolution of 0.5 $-$ 2.5 kpc.
Using broadband UV/optical/IR population synthesis, they determined the 
FUV optical depth within these galaxies, and compared to 
tracers of the interstellar gas.  Assuming
a slab geometry (i.e., a uniformly mixed
distribution of stars and dust), they derive the relationship
$\tau$$_{\rm FUV}$ = (2.226 $\pm$ 0.040) + (0.058 $\pm$ 0.001) $\times$
$\Sigma$$_{\rm H}$, where 
$\Sigma$$_{\rm H}$ is in M$_{\sun}$~pc$^{-2}$.
We combine this relationship with the 
spatially-resolved
Schmidt-Kennicutt law, and plot it on our FUV $-$ [24]
vs.\ SFR plot in Figure 17 (magenta
dotted curve) along with our best linear fit to the data
(blue solid curve).
The combination of the 
\citet{boquien13} attenuation model and the Schmidt-Kennicutt
law provides a fairly good match to our data.

At the high SFR end of the plots in Figures 16 and 17, 
the datapoints appear to have higher attenuations at H$\alpha$
and 
FUV than expected based on a linear fit 
(blue solid line),
the 
M51-derived expectation (Figure 16; dashed cyan curves), or
the \citet{boquien13} relation (Figure 17; dotted magenta curves).  
In other words, for the highest SFR clumps the SFRs are lower than
expected based on these relationships.
There has been some discussion recently in the literature 
suggesting that the Schmidt-Kennicutt law in starburst galaxies may
differ from that in spirals \citep{daddi10, boquien11}, with
significantly higher SFRs per gas surface density in the starbursts
(i.e., a `starburst mode' with 
higher star formation efficiencies).
In our highest SFR clumps, however, the SFR per dust attenuation
is lower than expected based on the correlations seen at lower
SFRs.   This is the opposite of what is expected if the efficiency
of star formation is higher in these regions, unless the 
dust attenuation to gas column density relation also differs in
these regions.
If large quantities of gas are flowing into these regions
but are not yet engaged in star formation, the attenuation may
be high relative to the SFR.
High spatial resolution CO observations of some of these high SFR
regions, particularly the hinge clumps in Arp 240 and Arp 256,
would be useful in testing this scenario by
determining the 
$\Sigma$$_{SFR}$$-$ N$_{\rm H}$ 
relation directly in these regions.

\section{Summary}

We extracted fluxes for clumps of star formation within 46 pre-merger
interacting galaxies and 39 isolated spirals.
We find a flatter distribution
of SFRs for the clumps in the interacting galaxies (i.e., proportionally more
clumps at higher luminosities) than those in the spirals.
If our tracers of SFR are consistent from region to region and
are unbiased, 
this implies that there are more young stars, on average, per region
in the interacting galaxies (i.e., a higher density
of stars).  This means that either the efficiency of star formation
per region is higher in interacting
galaxies, or there is more interstellar gas per clump,
or both.  
On average, a larger fraction of the total flux of the
interacting galaxies is contained in the clumps, compared
to the spirals. 

With the exception of the overlap region in the Antennae, 
the highest SFR clumps in our sample are 
located near the base of tidal features in interacting
galaxies.  Strong star formation 
may be triggered in these regions by converging gas flows along 
narrow caustics \citep{struck12}.
Published HST images 
reveal 
unusually large and luminous star clusters at the heart of the most
luminous clumps in our sample \citep{smith14}.

For the highest SFR regions, the implied H$\alpha$-to-UV dust attenuation
is consistent with the Calzetti starburst law, assuming an approximately
constant SFR over $\sim$100 Myrs.  This suggests that the young
stars in these regions are surrounded by a clumpy dusty shell of 
gas, with moderate star/ionized gas differential dust attenuation.
For the lower luminosity clumps, either the SFR has faded in the last
$\sim$10 $-$ 50 Myrs or there is enhanced differential attenuation
between the FUV-emitting stars and the ionized gas.

The inferred dust attenuation of our clumps increases with increasing
SFR.  This is consistent with the Schmidt-Kennicutt relation
between SFR and gas surface density, with regions with higher
gas content having higher SFRs.
We do not see evidence for a `starburst' mode of star formation
(i.e., high star formation efficiency) in our highest SFR clumps, 
if the dust attenuation to gas column density relations for these regions
are similar to those of the other clumps in the sample.

\acknowledgements

This research was supported by 
National Science Foundation
Extragalactic Astronomy grant AST-1311935.
We thank the anonymous referee for very helpful comments that greatly improved
this paper.  
We also thank Joan Font and Artemi Camps-Fari\~na for help with the 
WHT observations. 
We also acknowledge Mark Giroux and Brad Peterson for helpful comments, 
and Ashton Morelock
for help with downloading and checking data.
This research has made use of the NASA/IPAC Extragalatic Database (NED), which is operated by the Jet Propulsion Laboratory, California Institute of Technology, 
under contract with NASA.
This work is based in part on observations made
with the Spitzer Space Telescope, which is operated by
the Jet Propulsion Laboratory (JPL), California Institute
of Technology under a contract with NASA.
This study also uses data from the NASA Galaxy
Evolution Explorer (GALEX), which was operated for NASA
by the California Institute of Technology under
NASA contract NAS5-98034.
The William Herschel 4.2m Telescope is operated on the island of La Palma 
by the Isaac Newton Group in the Spanish Observatorio del Roque de 
los Muchachos of the Instituto de Astrofísica de Canarias.

\appendix{\bf Appendix: Using Spitzer Colors to Identify Interlopers}

In the absence of optical spectra or detections in H$\alpha$ maps,
a useful way of determining whether 8 $\mu$m-selected sources are 
foreground stars, background quasars, or star forming regions 
is to use the Spitzer broadband colors.
The 
Spitzer [3.6 $\mu$m] $-$ 
[4.5 $\mu$m] vs.\ [3.6 $\mu$m] $-$ [ 8.0 $\mu$m] 
color-color diagram is especially useful for
this task \citep{smith05, lapham13}. 
This color-color plot is a particularly useful diagnostic for our study,
since all of the galaxies in our sample have images in these Spitzer bands.
The global 3.6 $\mu$m
emission from galaxies is generally assumed to be dominated by light
from the older stellar population/underlying stellar mass 
while interstellar contributions
dominate at wavelengths of 5.8 $\mu$m and longer 
(e.g., \citealp{helou04}).
In the 3.6 $\mu$m $-$ 8.0 $\mu$m wavelength range, 
stars have spectral energy distributions (SEDs) that drop with wavelength,
giving [3.6] $-$ [4.5] and [3.6] $-$ [8.0] colors close to zero 
(e.g., \citealp{whitney04}).
In contrast, star forming regions have SEDs that drop from 3.6 $\mu$m
to 4.5 $\mu$m, and then increase at longer wavelengths
\citep{smith05, smith08, smith14,
lapham13, 
higdon14}. 
Quasars typically have flat SEDs in this wavelength range, giving
them redder [3.6] $-$ [4.5] colors than both stars and most 
star forming regions \citep{hatz05, smith05}.

In Figures 18 and 19, we 
display [3.6] $-$ 
[4.5] vs.\ [3.6] $-$ [ 8.0] 
color-color plots for our 1.0 kpc clump sample, with the 
clumps color-coded according to their classification.
Figure 18 plots the colors of the clumps in and near
the interacting galaxies, while
Figure 19 gives the clumps for the spirals.
The corresponding plots for the 2.5 kpc radii clumps are similar and thus
not shown.
The left panels of Figures 18 and 19 show the disk clumps 
(black open squares for the interacting galaxies and 
filled blue triangles for the spirals) and the tidal clumps in
the interacting galaxies (magenta
open diamonds). 
The right panels of Figures 18 and 19 
give the location of the nuclei (red open diamonds) and `off' clumps
(small green open squares).
For comparison,
in Figures 18 and 19
we also include Spitzer colors for H~II regions
in the Small and Large Magellanic Clouds from \citet{lawton10},
with the blue crosses being regions in the LMC and the cyan plus signs
being SMC regions. 
In Figures 18 and 19, we mark the expected location of
stars by blue dotted lines \citep{whitney04}, and that of quasars
by black dashed lines \citep{hatz05}.
Note that many of the objects classified as `off' sources lie in these
marked areas, along with a few of the galactic nuclei.
Proportionally, only a few of the clumps classified as `disk' or `tidal' 
are found in these areas.

Note that for the disk clumps in both the interacting sample and the spiral
galaxies, there is a trend such that the regions with the reddest
[3.6] $-$ [8.0] colors are also somewhat redder on average in [3.6] $-$ [4.5]
(see left panels in Figures 18 and 19).
This is likely due to interstellar contributions at 4.5 $\mu$m
(e.g., \citealp{lapham13, smith14}).
For spiral galaxies as a whole,
both the 3.6 $\mu$m and 4.5 $\mu$m Spitzer bands are generally assumed to be
dominated by starlight, as the 
global [3.6] $-$ [4.5] colors of spirals are usually close to zero 
(within 0.1 magnitudes) (e.g., \citealp{pahre04, smith07}).
However,
within interacting galaxies localized knots
of intense star formation
sometimes have large excesses above
the stellar continuum in the 3.6 $\mu$m and 4.5 $\mu$m bands, particularly
in the 4.5 $\mu$m filter
\citep{smith08, zhang10, boquien10}.

In Figures 18 and 19, the SMC regions 
are redder in [3.6] $-$ [4.5] 
for the same [3.6] $-$ [8.0] color compared to 
the regions within the interacting galaxies.
This may be a consequence of lower metallicity.
Such an offset is also seen in the global colors of 
low metallicity dwarfs compared
to higher metallicity galaxies \citep{smith09}.

Some of our `off' sources are found in the same part of the
color-color diagram as the `disk' and `tail' sources.  
The nature of these sources is uncertain.
They may be either
star forming regions in the outskirts of our target galaxies,
outside of our isophote limits,
or
they may be background galaxies.  
Some of the `off' sources 
lie in the same section of the color-color diagram
as the SMC regions, as do some of the
clumps classified as `disk' and `tidal'.
If they were not detected in H$\alpha$, 
the identication of these sources 
is ambiguous;
they
could be either low metallicity star forming regions in the outskirts
of the
target galaxy or
background objects.

\twocolumn

\begin{figure}
\plotone{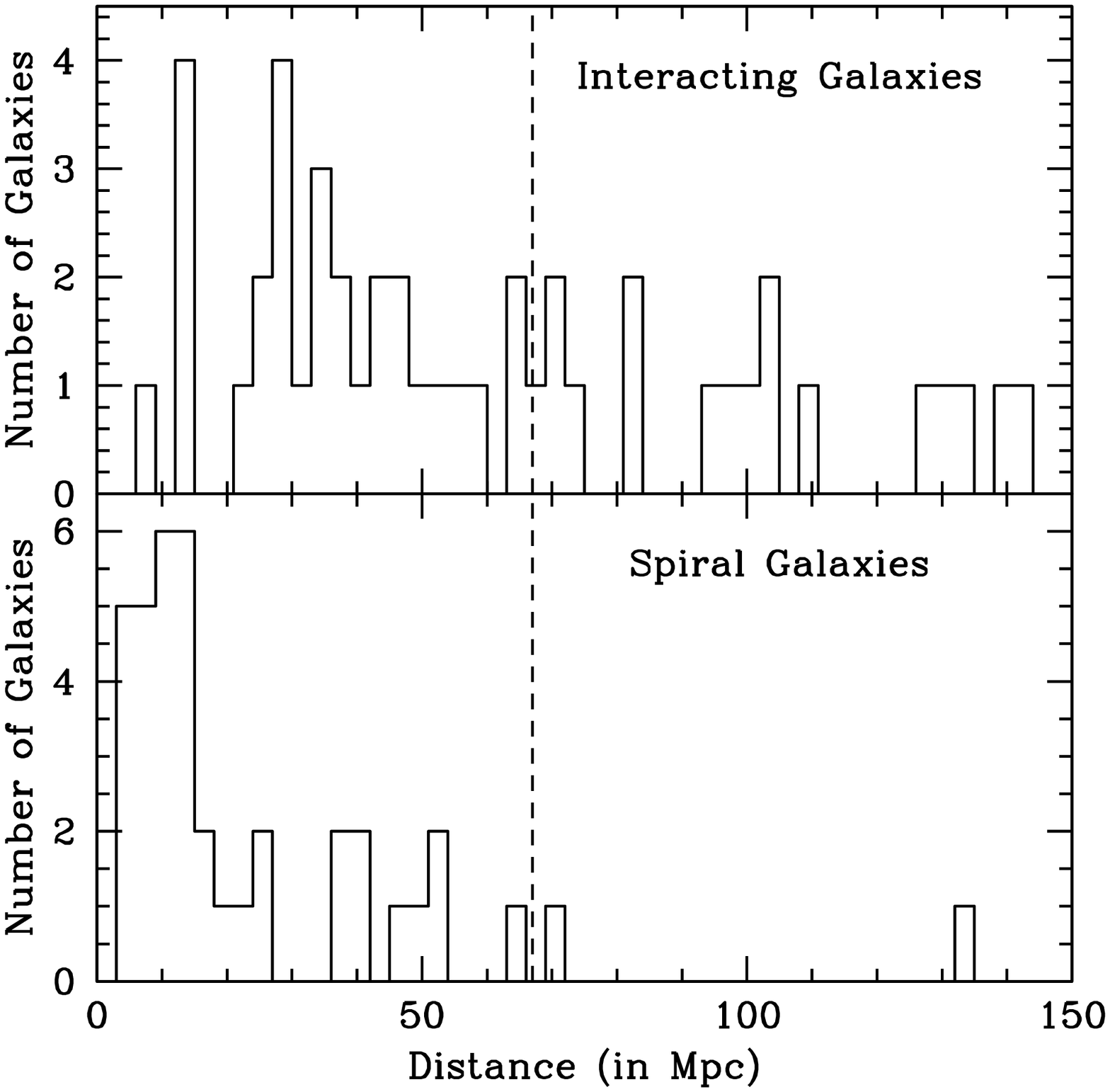}
\caption{
Histograms of the distances to the 
interacting galaxies (top panel) and the spiral galaxies
(bottom panel).
The dashed line at 67 Mpc shows the limit for the 1.0 kpc clump sample
(see Section 4.1).
}
\end{figure}

\begin{figure}
\plotone{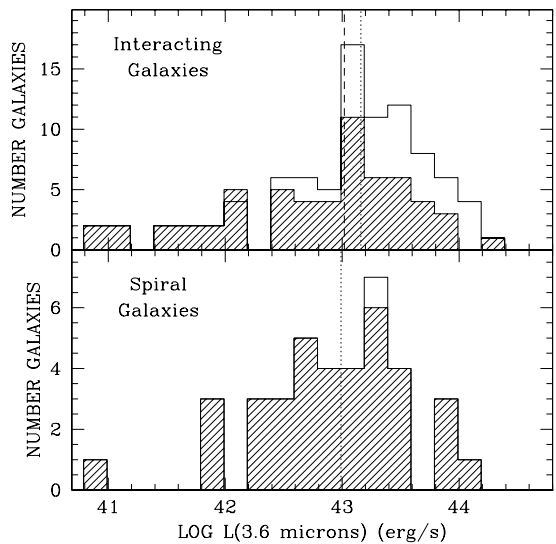}
\caption{
Histograms of the total 3.6 $\mu$m luminosities ($\nu$L$_{\nu}$)
for the individual galaxies in the interacting galaxy pairs
(top panel) and the spiral galaxies
(bottom panel).
The hatched histogram marks the galaxies in the 1.0 kpc clump
sample (i.e., those with distances less than
67 Mpc).
In the top panel,
the dotted line marks the median value for the entire interacting
galaxy sample 
(log L$_{3.6}$ = 43.16),
while the dashed line shows the median for the interacting galaxies 
with distances less than 67 Mpc
(log L$_{3.6}$ = 43.02).
In the bottom panel, the dotted line is the median for both sets of
spirals  
(log L$_{3.6}$ = 42.99).
Note that no Spitzer 3.6 $\mu$m data is available for
one of the more distant spiral galaxies, NGC 2857.
}
\end{figure}

\newpage
\onecolumn

\begin{figure}
\includegraphics[width=420pt]{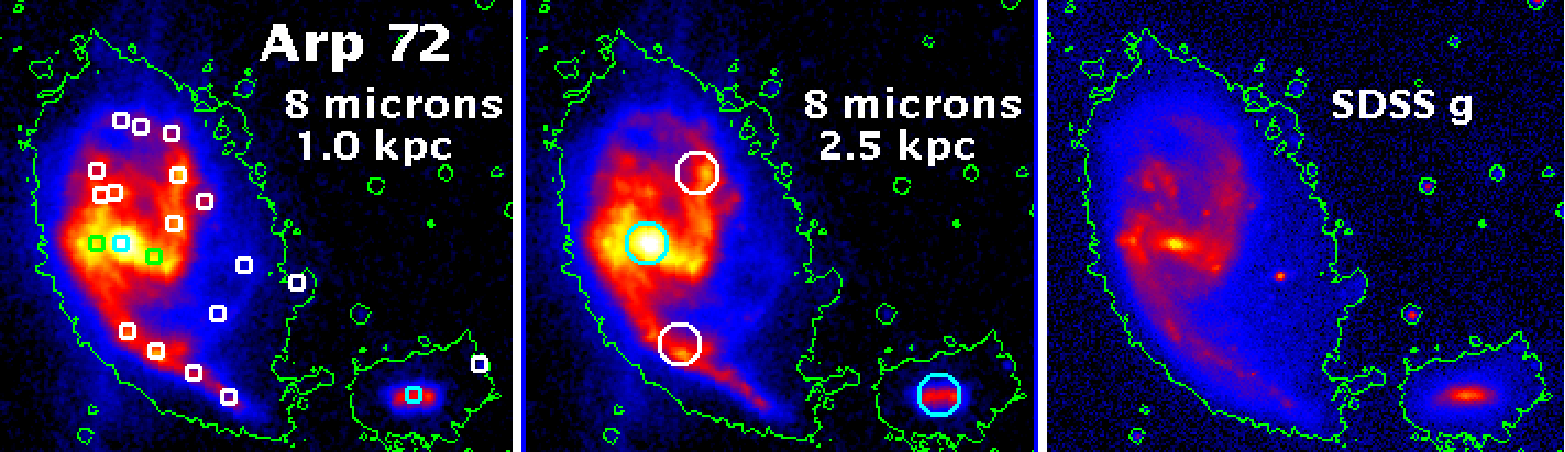}
\includegraphics[width=420pt]{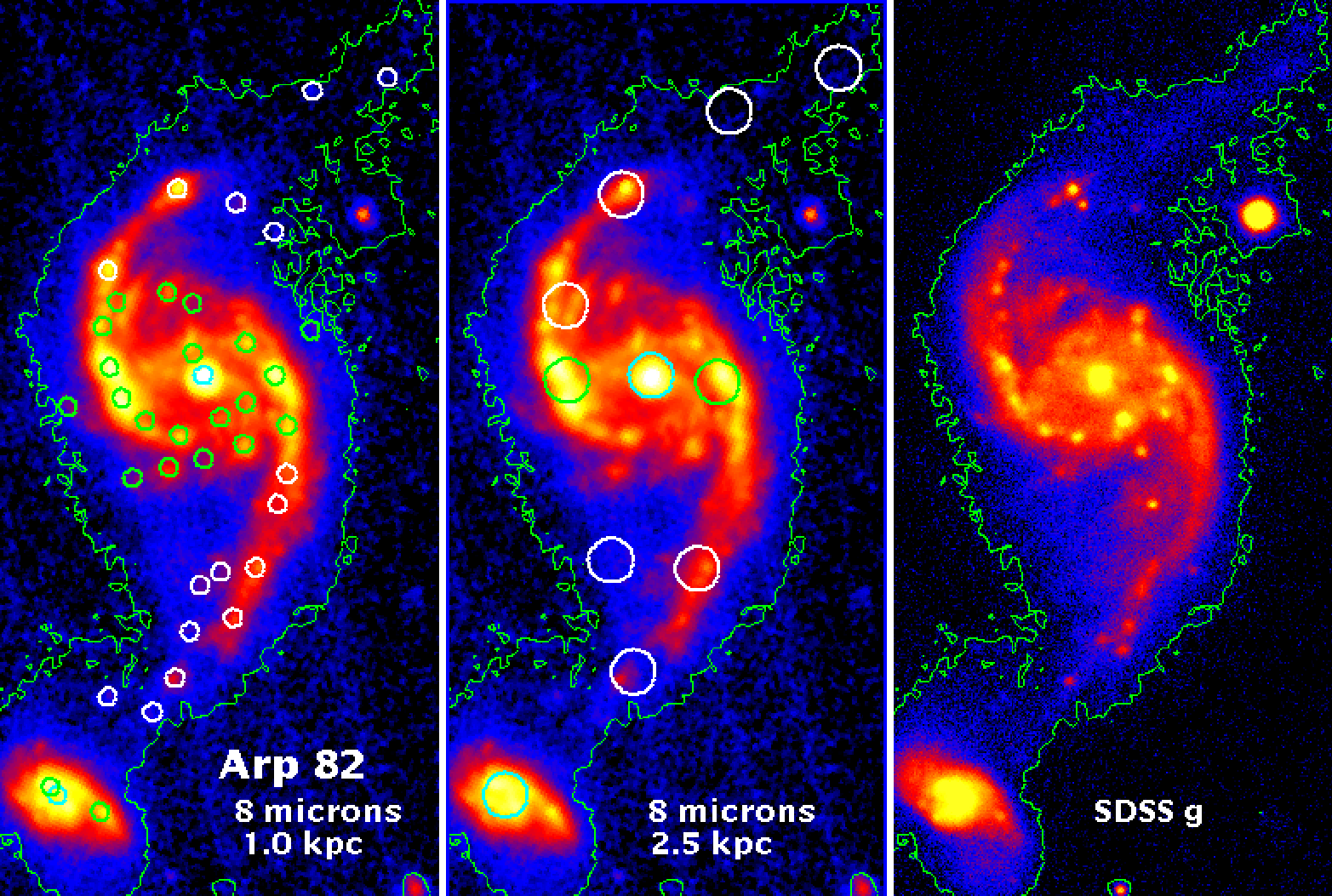}
\includegraphics[width=420pt]{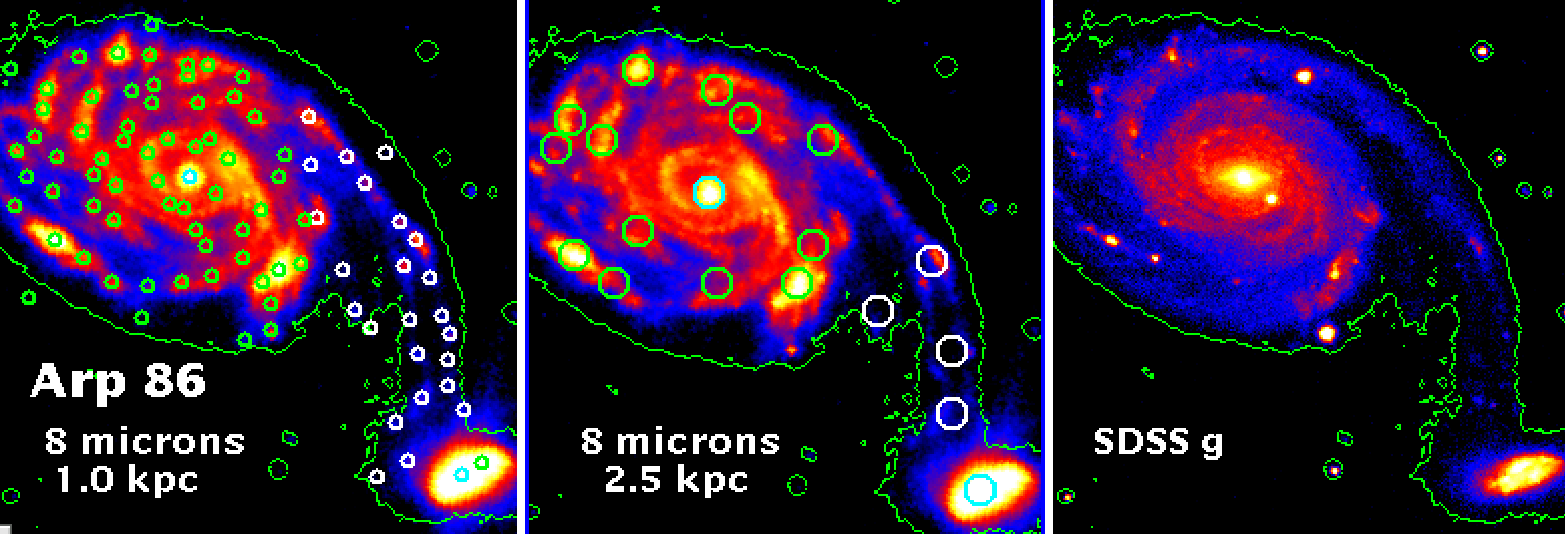}
\caption{
Three of the interacting galaxies in the sample, 
with the disk (green circles), tidal (white circles),
and nuclear (cyan circles) marked.
The left and middle panels
display the 1.0 kpc and 2.5 kpc clump samples, respectively, superimposed
on the unsmoothed Spitzer 8 $\mu$m image.
The right panels show the SDSS g image.  The green contours are the SDSS
g isophote of 24.58 mag/arcsec$^2$.
The top row are the images for Arp 72; the second row display the Arp 82
images, while the bottom row shows the Arp 86 images.
North is up and east to the left in these pictures.
}
\end{figure}

\twocolumn

\begin{figure}
\plotone{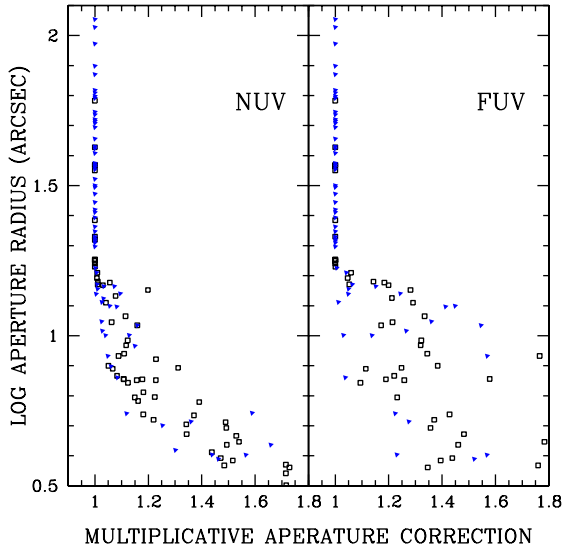}
\caption{
Aperture corrections for the GALEX 
photometry, as a function of aperture
radius.  The NUV values are on the left; the FUV on the right.
This plot includes images from both the interacting (black open squares)
and spiral (blue filled triangles) samples,
and measurements for both the 1.0 kpc radius apertures and the
2.5 kpc radius apertures.
See text for details.
}
\end{figure}

\begin{figure}
\plotone{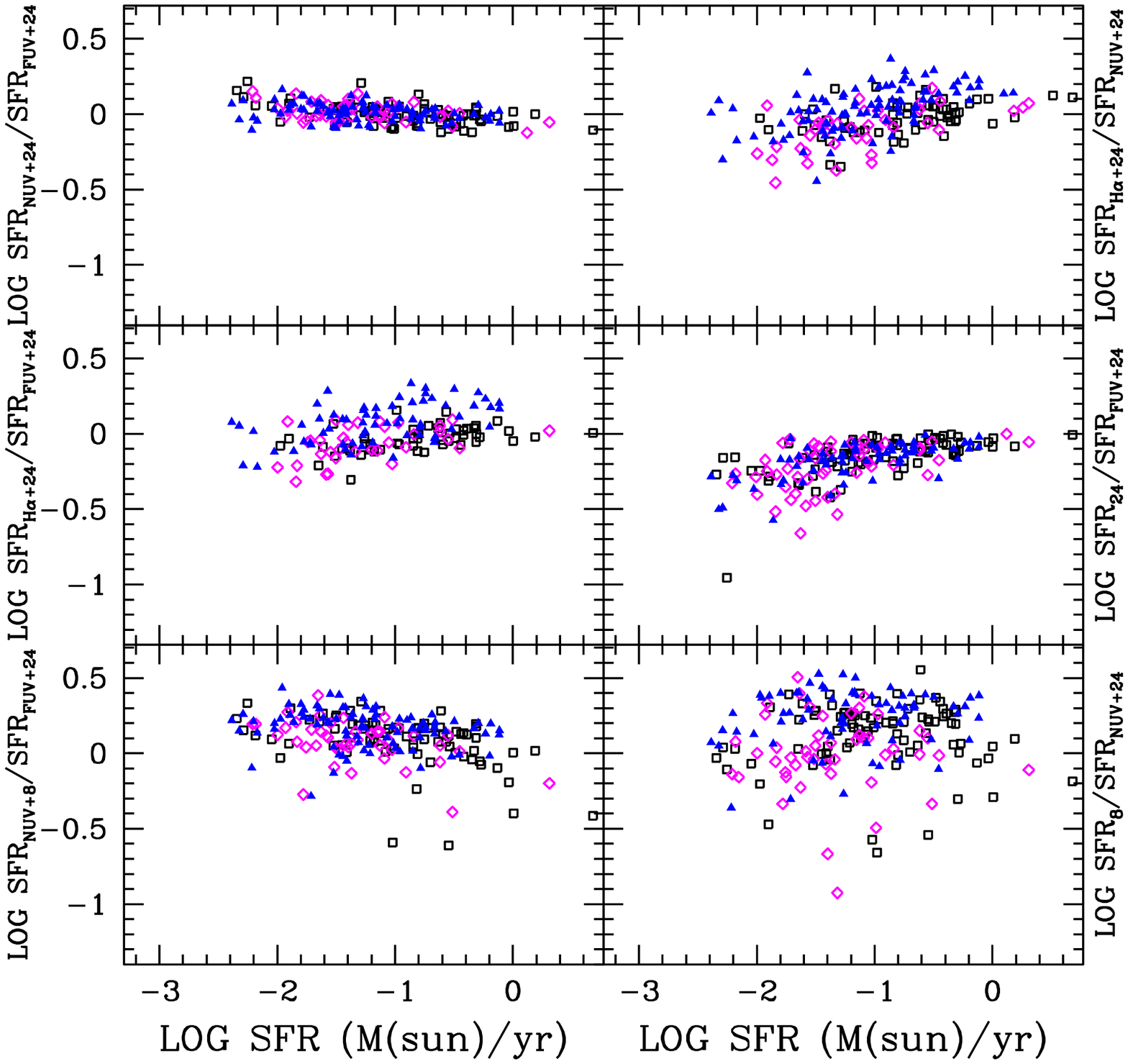}
\caption{For the 2.5 kpc aperture radius clump samples,
this plot provides 
a comparison of the SFR determined from the FUV and 24 $\mu$m 
luminosities (x-axis) against various ratios of different determinations
of the SFRs.
Disk clumps 
in the interacting galaxies are shown as black open squares,
clumps in the disks of the spirals are shown as blue filled triangles,
and the tidal clumps in the interacting galaxies are shown as open
magenta diamonds. 
}
\end{figure}

\begin{figure}
\plotone{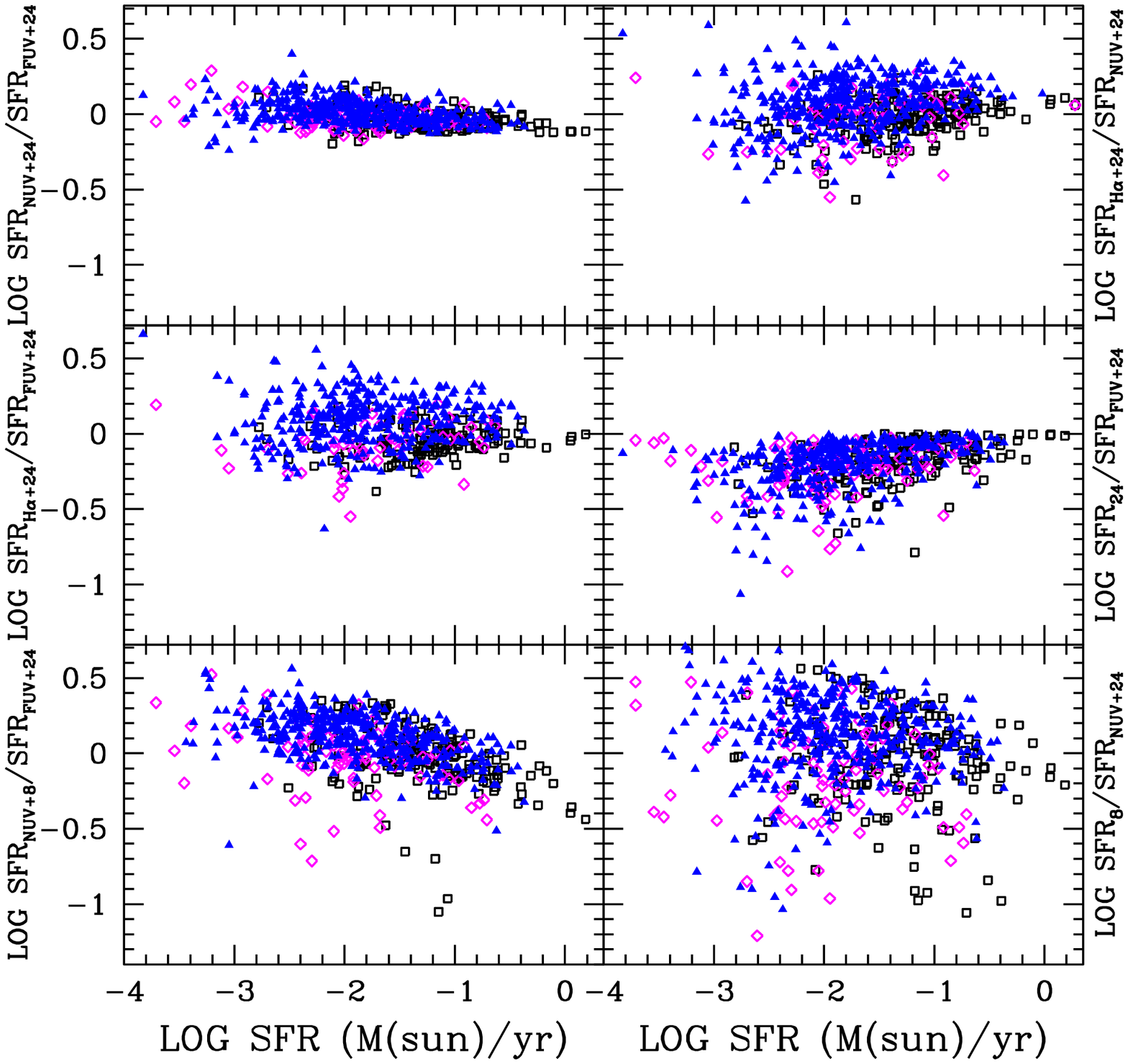}
\caption{For the 1.0 kpc aperture radius clump samples,
this plot provides 
a comparison of the SFR determined from a combination of
the FUV and the 24 $\mu$m 
luminosities (x-axis) against various ratios of different determinations
of the SFRs.
Disk clumps 
in the interacting galaxies are shown as black open squares,
clumps in the disks of the spirals are shown as blue filled triangles,
and the tidal clumps in the interacting galaxies are shown as open
magenta diamonds. 
}
\end{figure}

\begin{figure}
\plotone{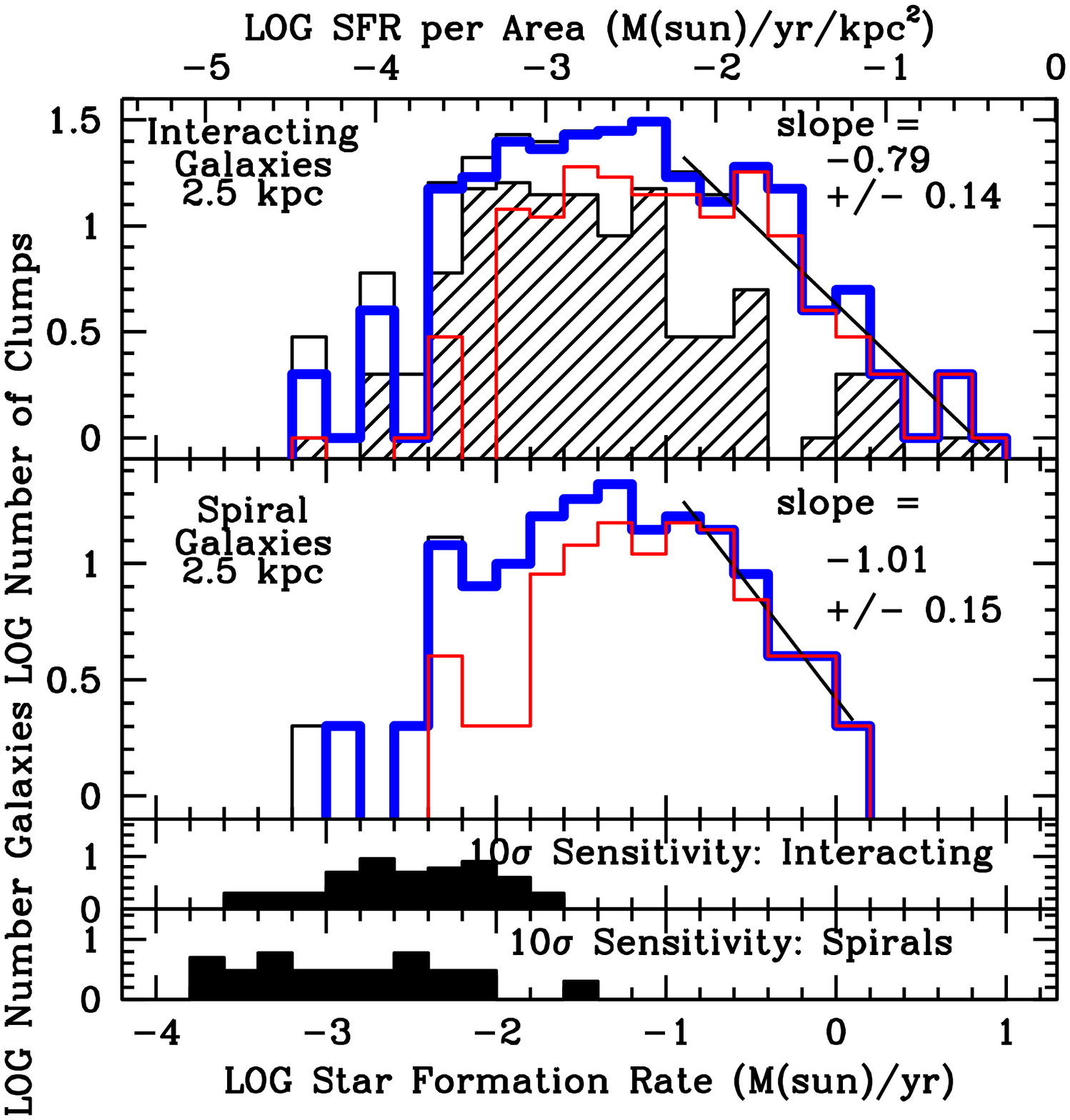}
\caption{Histograms of the SFRs
within the 2.5 kpc sample
for the clumps in the interacting
galaxies (top panel), 
and the disks of the normal spiral galaxies
(second panel).
Along the top axis of this figure, we have converted the SFRs into 
SFR per area,
by dividing by the area per aperture. 
In the top panel, the tidal clumps are shown with hatch marks.
The red histogram identifies clumps detected in H$\alpha$, while
the blue includes both clumps detected in H$\alpha$ and clumps
that lie outside of the quasar and stellar regions marked in the
Spitzer color-color diagrams (see Appendix).
The lines are the best-fit lines to the blue histogram
above log SFR $>$ $-$0.8.
Histograms of the theoretical
10$\sigma$ point source sensitivities of the smoothed 8 $\mu$m images 
of the interacting and spiral galaxies are presented in the 
third and bottom panels, respectively.
In the two bottom panels, the y-axis is the logarithm of the number
of galaxies at each sensitivity.
}
\end{figure}

\begin{figure}
\plotone{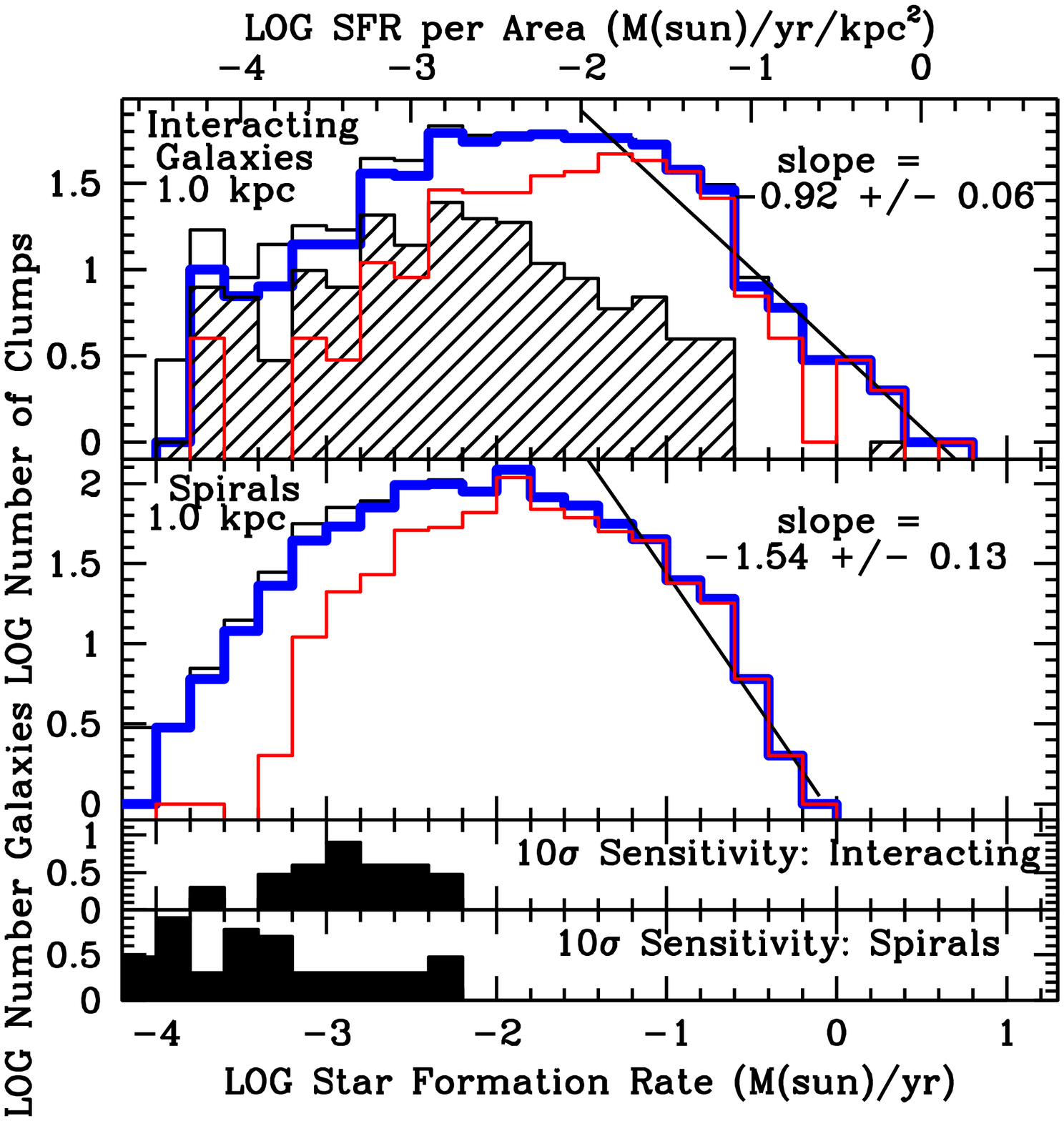}
\caption{Histograms of the SFRs
within the 1.0 kpc sample
for the clumps in the interacting
galaxies (top panel), 
and the disks of the normal spiral galaxies
(second panel).
Along the top axis of this plot, we have converted the SFRs into SFR per area,
by dividing by the area per aperture. 
In the top panel, the tidal clumps are shown with hatch marks.
The red histogram identifies clumps detected in H$\alpha$, while
the blue includes both clumps detected in H$\alpha$ and clumps
that lie outside of the quasar and stellar regions marked in the
Spitzer color-color diagrams (see Appendix).
The lines are the best-fit lines to the blue histogram
above log SFR $>$ $-$1.6.
Histograms of the theoretical
10$\sigma$ point source sensitivities of the smoothed 8 $\mu$m images 
of the interacting and spiral galaxies are presented in the 
third and bottom panels, respectively.
In the two bottom panels, the y-axis is the logarithm of the number
of galaxies at each sensitivity.
}
\end{figure}

\begin{figure}
\plotone{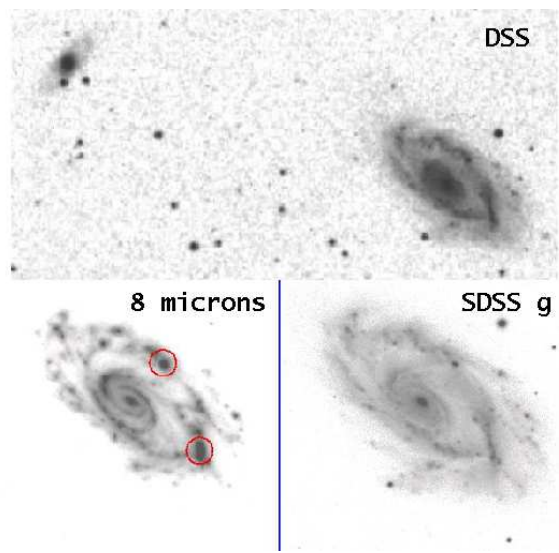}
\caption{Top: the optical Digitized Sky Survey (DSS)
image of NGC 3646 (larger galaxy)
and its companion NGC 3649. North is up and east to the left.
The field of view is 11$'$ $\times$ 5\farcm5.  
Bottom left and right, respectively:
the unsmoothed Spitzer 8 $\mu$m and SDSS g images of NGC 3646.
The field of view is 3\farcm4 $\times$ 3\farcm4.
The two highest SFR clumps are circled
in red in the 8 $\mu$m image.  
}
\end{figure}

\begin{figure}
\plotone{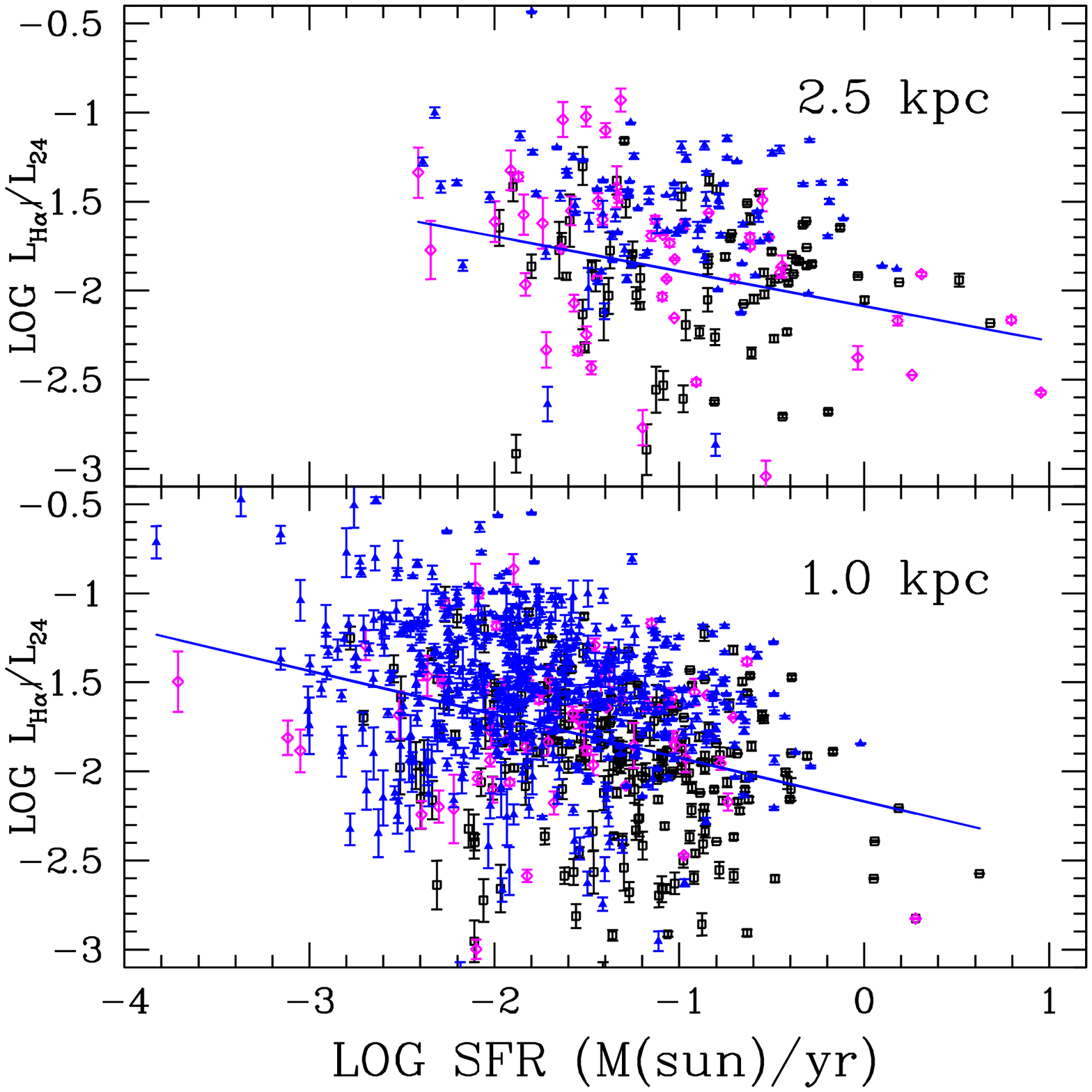}
\caption{Plots of 
L$_{H\alpha}$/L$_{24}$ vs.\ SFR for the 2.5 kpc clump sample (top panel)
and the 1.0 kpc clump sample (bottom panel).
The open black squares are the disk clumps in the interacting galaxies,
while the magenta open diamonds are the tidal clumps.
The small blue filled triangles mark clumps in the disks of the
spirals.
The solid blue line displays the best linear fit to the data.
}
\end{figure}

\begin{figure}
\plotone{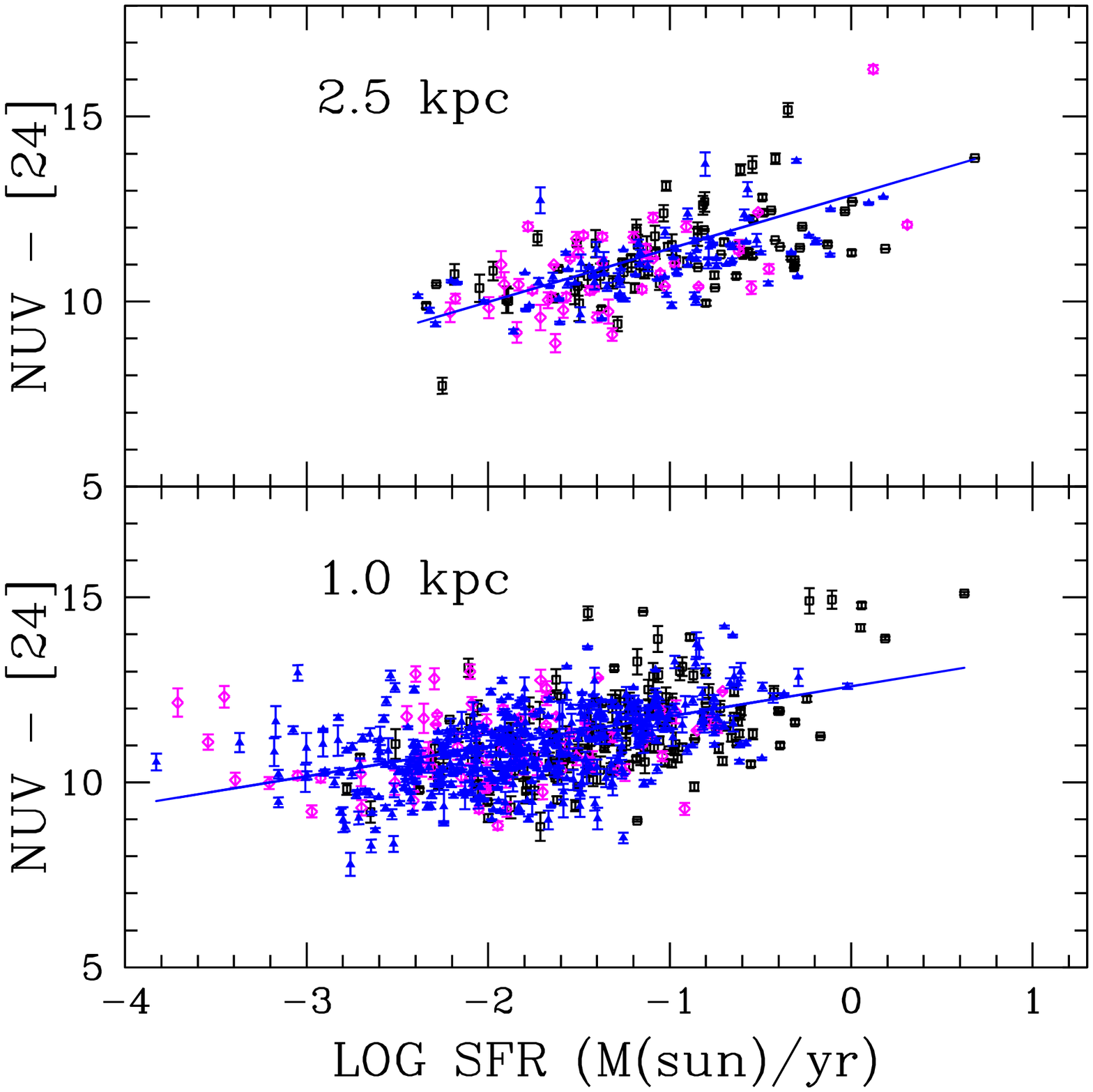}
\caption{Plots of 
NUV - [24] vs.\ SFR for the 2.5 kpc clump sample (top panel)
and the 1.0 kpc clump sample (bottom panel).
The open black squares are the disk clumps in the interacting galaxies,
while the magenta open diamonds are the tidal clumps.
The small blue filled triangles mark clumps in the disks of the
spirals.
The solid blue line displays the best linear fit to the data.
}
\end{figure}

\begin{figure}
\plotone{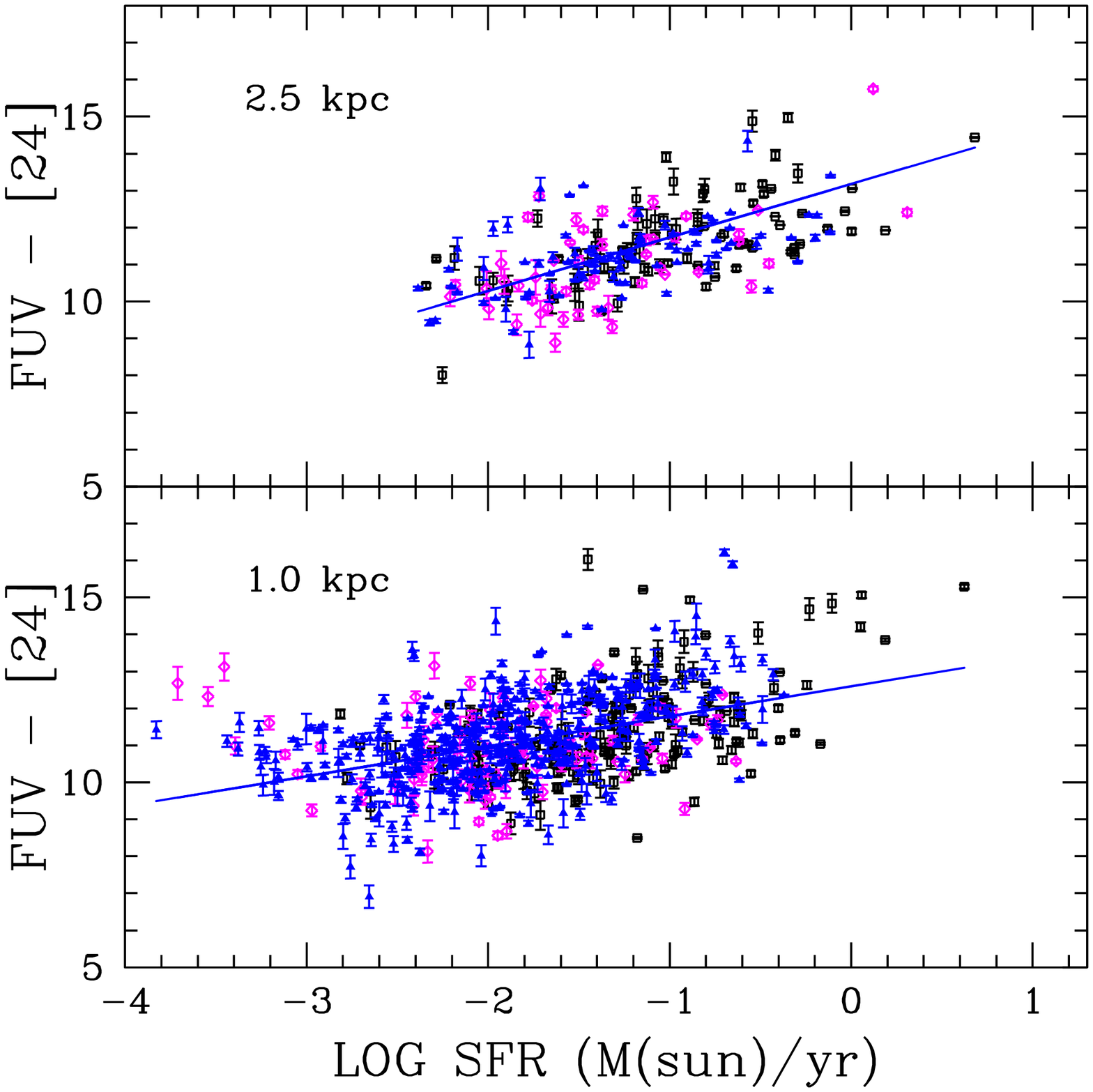}
\caption{Plots of 
FUV - [24] vs.\ SFR for the 2.5 kpc clump sample (top panel)
and the 1.0 kpc clump sample (bottom panel).
The open black squares are the disk clumps in the interacting galaxies,
while the magenta open diamonds are the tidal clumps.
The small blue filled triangles mark clumps in the disks of the
spirals.
The solid blue line displays the best linear fit to the data.
}
\end{figure}

\begin{figure}
\plotone{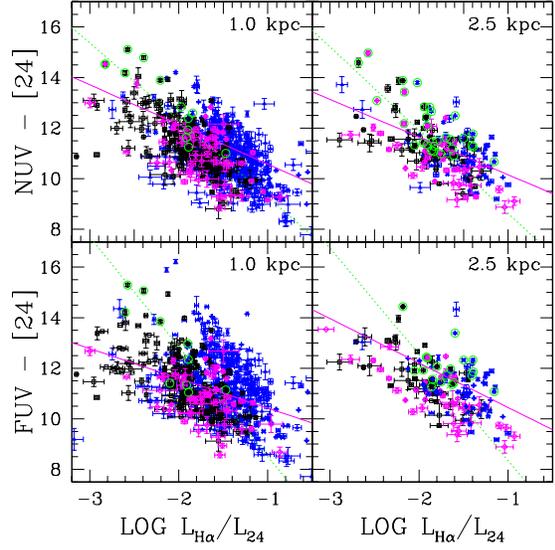}
\caption{Plots of 
NUV - [24] vs.\ log(L$_{H\alpha}$/L$_{24}$ (top row) 
and FUV - [24] vs.\ L$_{H\alpha}$/L$_{24}$ 
(bottom row)
for the 2.5 kpc clump sample (right panels)
and the 1.0 kpc clump sample (left panels).
The open black squares are the disk clumps in the interacting galaxies,
while the magenta open diamonds are the tidal clumps.
The small blue filled triangles mark clumps in the disks of the
spirals.
The clumps circled in green have SFRs $>$ 0.4 M$_{\sun}$~yr$^{-1}$.
The magenta solid curves gives
the best linear fits to all of the data, while the dotted green
lines are the best fits to the clumps with  
SFRs $>$ 0.4 M$_{\sun}$~yr$^{-1}$.
}
\end{figure}

\begin{figure}
\plotone{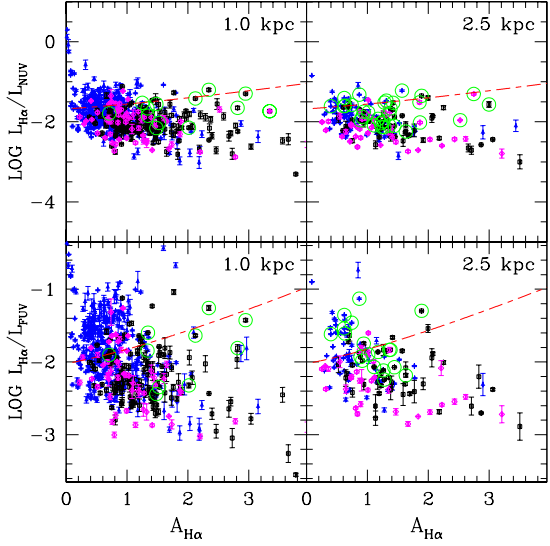}
\caption{Plots of 
log L$_{H\alpha}$/L$_{\rm NUV}$ vs. A$_{H\alpha}$
(top panels) and 
log L$_{H\alpha}$/L$_{\rm FUV}$ vs. A$_{H\alpha}$
(bottom panels)
for the 
1.0 kpc radii aperture (left panels)
and the 2.5 kpc radii aperture clumps (right panels).
The open black squares are the disk clumps in the interacting galaxies,
while the magenta open diamonds are the tidal clumps.
The small blue filled triangles mark clumps in the disks of the
spirals.
The clumps circled in green have SFRs $>$ 0.4 M$_{\sun}$~yr$^{-1}$.
The red dashed lines show the expected relationship for the Calzetti
dust attenuation law, combined with the NUV and FUV attenuations relative
to the 24 $\mu$m emission implied by the assumed SFR laws (see text).
}
\end{figure}

\begin{figure}
\plotone{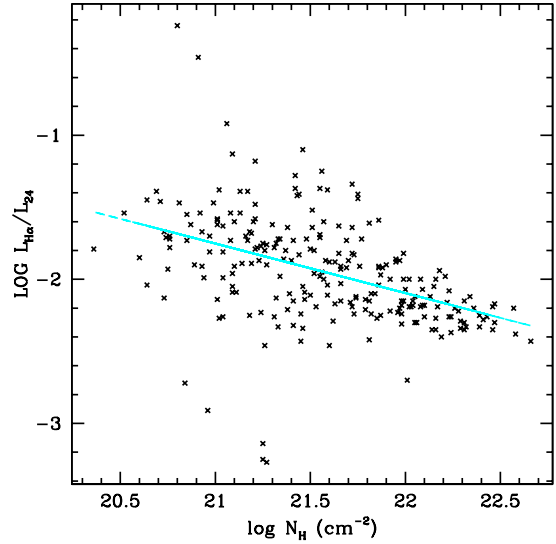}
\caption{
Plot of 
log L$_{H\alpha}$/L$_{24}$ vs.\ N$_{\rm H}$ 
within 500 pc regions
in M51 using data from \citet{kennicutt07}.
Our best-fit straight line to the data is 
plotted in cyan.
}
\end{figure}

\begin{figure}
\plotone{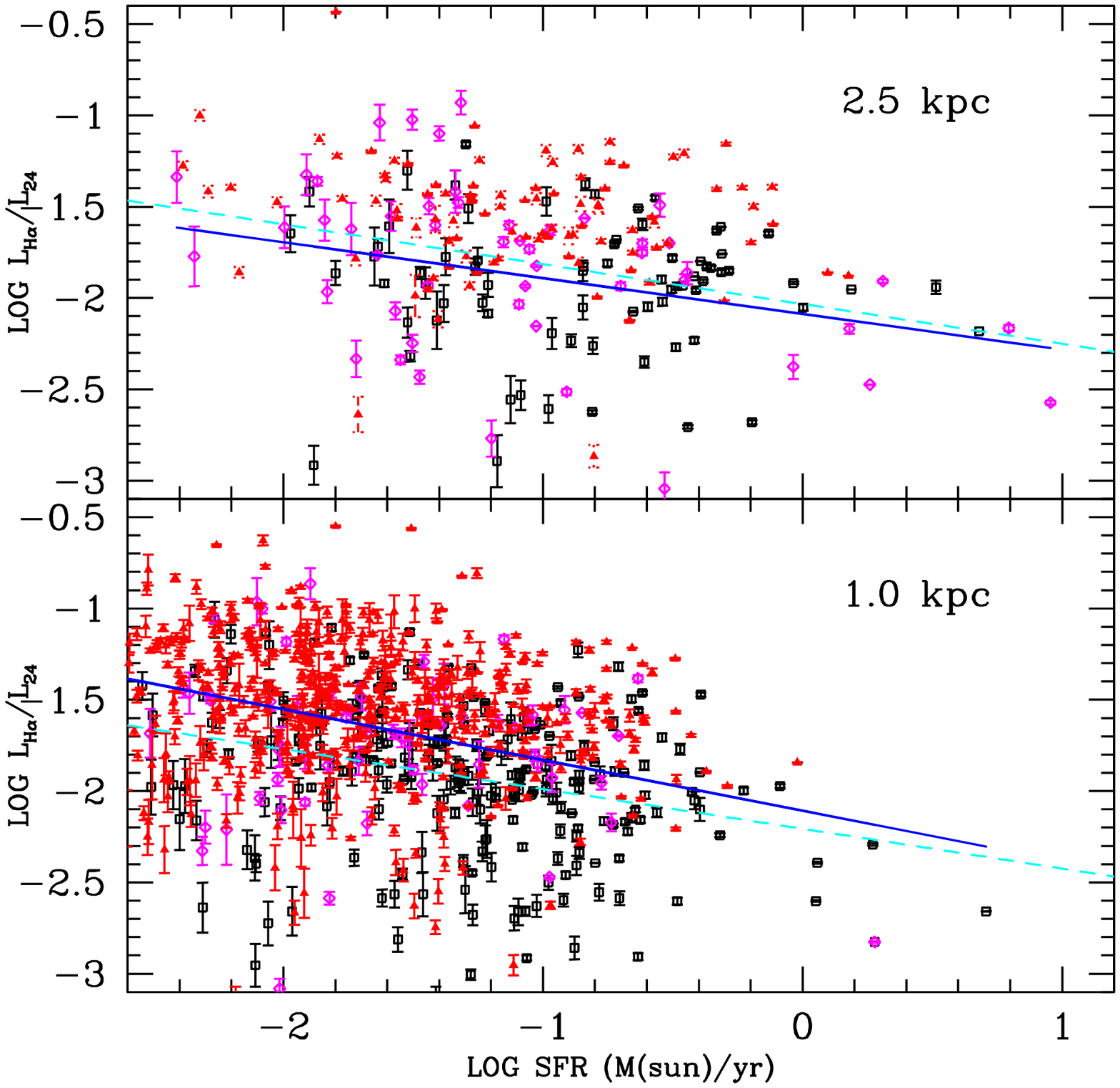}
\caption{Plots of 
L$_{H\alpha}$/L$_{24}$ vs.\ SFR for the 2.5 kpc clump sample (top panel)
and the 1.0 kpc clump sample (bottom panel).
The open black squares are the disk clumps in the interacting galaxies,
while the magenta open diamonds are the tidal clumps.
The small red filled triangles mark clumps in the disks of the
spirals.
The solid blue line displays the best linear fit to the data.
The dashed cyan curve comes from the 
L$_{H\alpha}$/L$_{24}$ vs.\ N$_{\rm H}$ relationship for M51 shown
in Figure 15,  
along with the spatially-resolved
Schmidt-Kennicutt law for M51 found by \citet{kennicutt07}.
}
\end{figure}

\begin{figure}
\plotone{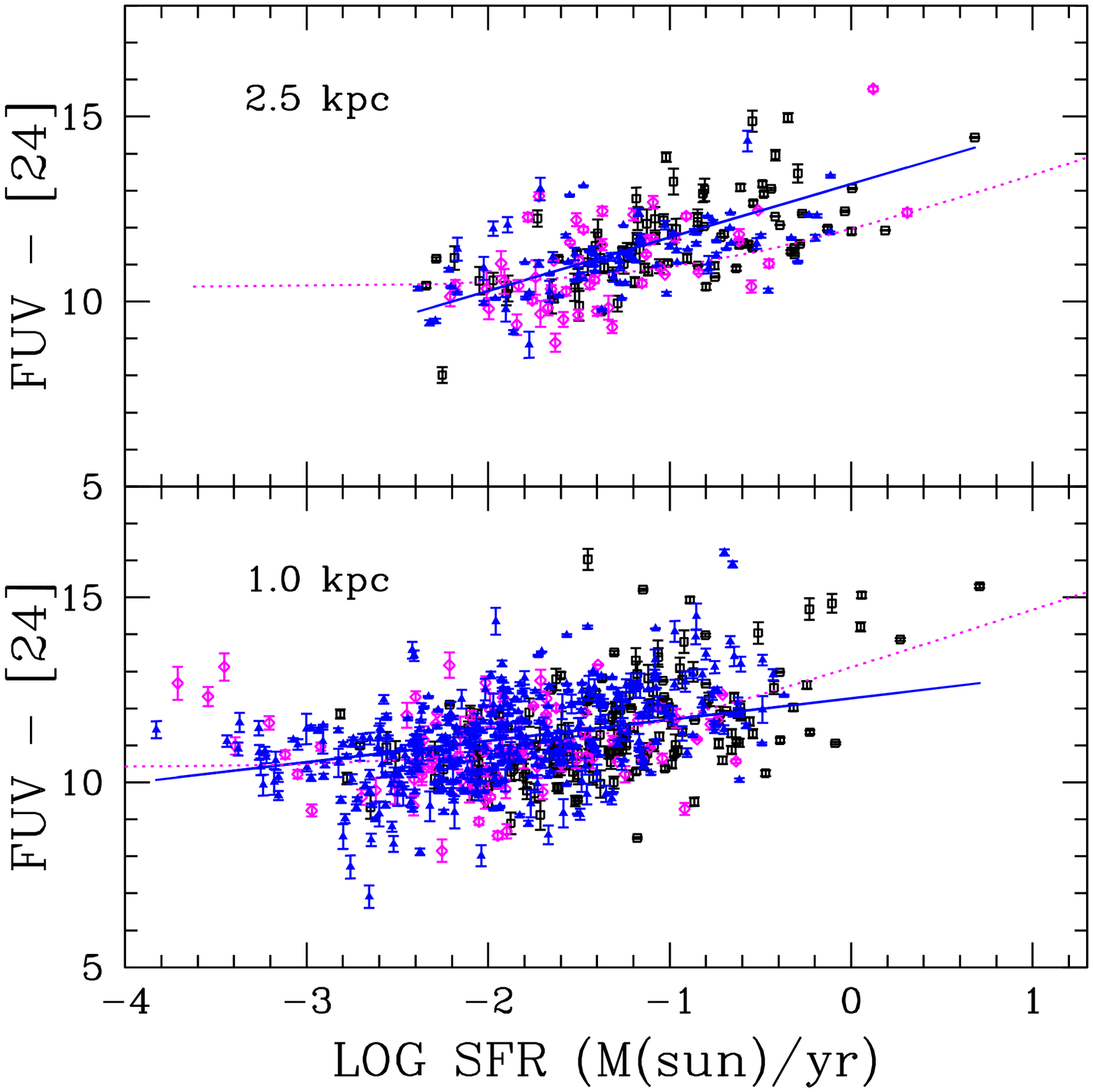}
\caption{Plots of 
FUV - [24] vs.\ SFR for the 2.5 kpc clump sample (top panel)
and the 1.0 kpc clump sample (bottom panel).
The open black squares are the disk clumps in the interacting galaxies,
while the magenta open diamonds are the tidal clumps.
The small blue filled triangles mark clumps in the disks of the
spirals.
The solid blue line displays the best linear fit to the data.
The dotted magenta curve gives the relation
derived using the \citet{boquien13} M33 $\tau$$_{\rm FUV}$$-$N$_{\rm H}$
relation 
and the spatially-resolved
Schmidt-Kennicutt law for M51 found by \citet{kennicutt07}.
}
\end{figure}

\begin{figure}
\plotone{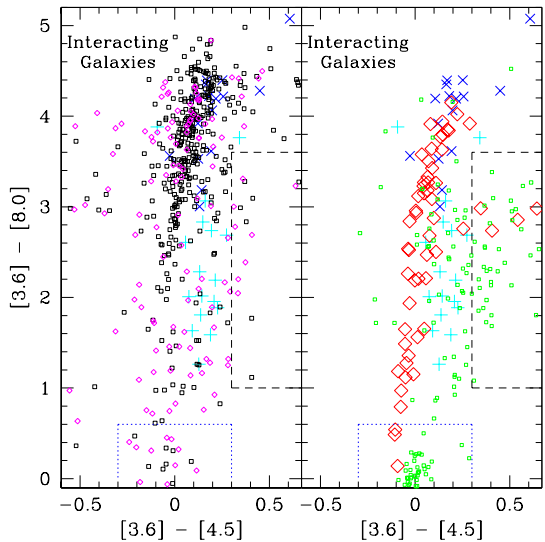}
\caption{Left: For the 1.0 kpc sample of clumps, a plot of the Spitzer 
[3.6] $-$ [4.5] vs.\ [3.6] $-$ [8]
colors
for the clumps in the disks
(small black squares) and in the tidal features
(small magenta diamonds) of the interacting galaxies.
Right: the same plot, but for 
the nuclei (red open diamonds) and `off' galaxy sources (small green open
squares).
In both plots, the 
LMC (blue crosses) and SMC (cyan plus signs)
regions 
from Lawton et al.\ (2010) are plotted.
For clarity, errorbars are omitted in these plots.   
These are generally about the size of
the data points or slightly larger.
The blue dotted rectangle approximately
marks the expected colors of foreground
stars, while quasars are typically found in the black dashed rectangle.
See the text for more details.
}
\end{figure}

\begin{figure}
\plotone{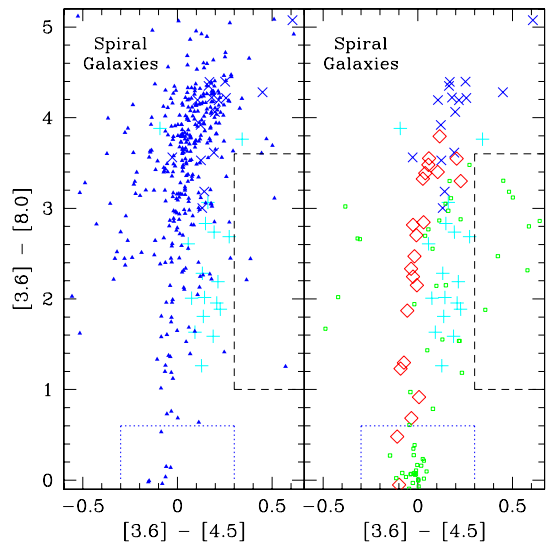}
\caption{Similar to Figure 18, but for the 1.0 kpc aperture
radii
clumps in the spiral galaxies.
The small blue filled triangles mark clumps in the disks of the
spirals.
Red open diamonds mark galactic nuclei, and 
'off' galaxy sources are shown by small green open
squares.
In both plots, the 
LMC (blue crosses) and SMC (cyan plus signs)
regions 
from Lawton et al.\ (2010) are plotted.
Errorbars are omitted in these plots.   
These are generally about the size of
the data points or slightly larger.
The blue dotted rectangle approximately
marks the expected colors of foreground
stars, while quasars are typically found in the black dashed rectangle.
See the text for more details.
}
\end{figure}

\clearpage
\input{sbt_table1.tex}
\clearpage

\clearpage
\input{sbt_table2.tex}
\clearpage

\end{document}

%% file: sbt_table1.tex
\begin{deluxetable}{rcrrrcrrrrccccccc}
\tabletypesize{\scriptsize}
\setlength{\tabcolsep}{0.05in}
\def\et#1#2#3{${#1}^{+#2}_{-#3}$}
\tablewidth{0pt}
\tablecaption{Interacting Galaxy Sample}
\tablehead{
\multicolumn{1}{c}{Arp} &
\multicolumn{1}{c}{Other} &
\multicolumn{1}{c}{D$^{\dagger}$}&
\multicolumn{1}{c}{log$^{\ddagger}$} &
\multicolumn{1}{c}{log}&
\multicolumn{1}{c}{H$\alpha$}
\\
\multicolumn{1}{c}{Name} &
\multicolumn{1}{c}{Name(s)} &
\multicolumn{1}{c}{(Mpc)} &
\multicolumn{1}{c}{L$_{FIR}$} &
\multicolumn{1}{c}{L$_{H\alpha}$} &
\multicolumn{1}{c}{Reference}
\\
\multicolumn{1}{c}{} &
\multicolumn{1}{c}{} &
\multicolumn{1}{c}{} &
\multicolumn{1}{c}{(L$_{\sun}$)} &
\multicolumn{1}{c}{$({\rm erg~s^{-1}})$}&
\multicolumn{1}{c}{} 
\\
}
\startdata

Arp 24  &  NGC  3445  &  33.1  &  9.6  &  41.0  &  This work: WHT  \\
Arp 34  &  NGC  4613/4/5  &  72.5  &  10.2  &  41.6  &  This work: WHT  \\
Arp 65  &  NGC  90/93  &  72.0  &  9.5  &  40.8  &  This work: WHT  \\
Arp 72  &  NGC  5994/6  &  53.4  &  10.3  &  41.6  &  This work: SARA  \\
Arp 82  &  NGC  2535/6  &  59.2  &  10.2  &  41.5  &  Hancock et al. (2007)  \\
Arp 84  &  NGC  5394/5  &  55.5  &  10.8  &  41.4  &  This work: WHT  \\
Arp 85  &  NGC  5194/5  &  12.1  &  10.3  &  41.6  &  Hoopes et al. (2001)  \\
Arp 86  &  NGC  7752/3  &  65.9  &  10.7  &  \ldots  &  \ldots  \\
Arp 87  &  NGC  3808  &  104.6  &  10.9  &  41.6  &  This work: WHT  \\
Arp 89  &  NGC  2648  &  31.8  &  9.0  &  40.2  &  This work: WHT  \\
Arp 91  &  NGC  5953/4  &  34.3  &  10.4  &  41.4  &  This work: WHT  \\
Arp 102  &  UGC 10814  &  104.7  &  9.7  &  40.4  &  This work: WHT  \\
Arp 104  &  NGC  5216/8  &  50.6  &  10.5  &  41.2  &  This work: WHT  \\
Arp 105  &  NGC  3561/UGC06224  &  126.2  &  11.0  &  41.1  &  This work: WHT  \\
Arp 107  &  UGC 5984  &  141.8  &  10.1  &  41.6  &  Smith et al. (2007)  \\
Arp 120  &  NGC  4435/8  &  14.0  &  9.2  &  40.1  &  This work: WHT  \\
Arp 178  &  NGC  5613/4/5  &  82.5  &  10.4  &  40.8  &  This work: WHT  \\
Arp 181  &  NGC  3212/5  &  132  &  10.6  &  40.8  &  This work: WHT  \\
Arp 188  &  UGC 10214  &  134.2  &  10.0  &  41.0  &  This work: WHT  \\
Arp 202  &  NGC  2719  &  47.6  &  9.8  &  41.4  &  This work: SARA  \\
Arp 205  &  NGC  3448/UGC6016  &  24.7  &  9.8  &  40.9  &  This work: WHT  \\
Arp 240  &  NGC  5257/8  &  101.7  &  11.3  &  42.2  &  Bushouse (1987)  \\
Arp 242  &  NGC  4676  &  98.2  &  10.7  &  41.4  &  This work: WHT  \\
Arp 244  &  NGC  4038/9  &  24.1  &  10.7  &  41.6  &  This work: WHT  \\
Arp 245  &  NGC  2992/3  &  34.0  &  10.4  &  40.8  &  This work: WHT  \\
Arp 253  &  UGC 173/4  &  28.8  &  8.7  &  \ldots  &  \ldots  \\
Arp 256  &     &  109.6  &  11.1  &  42.0  &  Bushouse (1987)  \\
Arp 261  &     &  28.7  &  9.3  &  40.8  &  This work: WHT  \\
Arp 269  &  NGC  4485/4490  &  8.5  &  9.8  &  40.2  &  This work: WHT  \\
Arp 270  &  NGC  3395/6  &  29.0  &  10.0  &  41.6  &  Zaragoza-Cardiel et al. (2013)  \\
Arp 271  &  NGC  5426/7  &  40.0  &  10.0  &  41.5  &  This work: SARA  \\
Arp 279  &  NGC  1253  &  22.6  &  9.4  &  41.0  &  This work: WHT  \\
Arp 280  &  NGC  3769  &  14.5  &  9.1  &  \ldots  &  \ldots  \\
Arp 282  &  NGC  169  &  64.9  &  10.1  &  \ldots  &  \ldots  \\
Arp 283  &  NGC  2798/9  &  29.6  &  10.5  &  41.2  &  SINGS  \\
Arp 284  &  NGC  7714/5  &  38.6  &  10.1  &  41.8  &  Smith et al. (1997)  \\
Arp 285  &  NGC  2854/6  &  44.4  &  10.1  &  41.2  &  This work: WHT  \\
Arp 290  &  IC 195/6  &  46.5  &  9.2  &  39.9  &  This work: WHT  \\
Arp 293  &  NGC  6285/6  &  82.2  &  11.1  &  \ldots  &  \ldots  \\
Arp 294  &  NGC  3786/8  &  43.6  &  \ldots  &  41.1  &  This work: WHT  \\
Arp 295  &     &  94.2  &  10.9  &  \ldots  &  \ldots  \\
Arp 297N  &  NGC  5753/5  &  139.3  &  11.1  &  40.9  &  This work: WHT  \\
Arp 297S  &  NGC  5752/4  &  70.2  &  10.3  &  \ldots  &  \ldots  \\
Arp 298  &  NGC  7469/IC5283  &  66.4  &  11.2  &  \ldots  &  \ldots  \\
NGC 2207/IC2163  &     &  38.0  &  10.7  &  41.7  &  Elmegreen et al. (2001)  \\
NGC 4567/8  &     &  13.9  &  9.9  &  40.9  &  Koopmann et al. (2001)  \\
\enddata
{
$^{\dagger}$From the NASA Extragalactic Database (NED), using
H$_0$ = 73 km/s/Mpc, with Virgo,
Great Attractor, and Shapley Supercluster infall models.
$^{\ddagger}$Total 42.4 $-$ 122.5 $\mu$m far-infrared luminosity from the Infrared
Astronomical Satellite (IRAS).
}
\end{deluxetable}
\clearpage

%% file: sbt_table2.tex
\begin{deluxetable}{rcrrrcrrrrccccccc}
\tabletypesize{\scriptsize}
\setlength{\tabcolsep}{0.05in}
\def\et#1#2#3{${#1}^{+#2}_{-#3}$}
\tablewidth{0pt}
\tablecaption{Spiral Galaxy Sample}
\tablehead{
\multicolumn{1}{c}{Name} &
\multicolumn{1}{c}{Type} &
\multicolumn{1}{c}{D$^{\dagger}$}&
\multicolumn{1}{c}{log$^{\ddagger}$} &
\multicolumn{1}{c}{log}&
\multicolumn{1}{c}{H$\alpha$}
\\
\multicolumn{1}{c}{Name} &
\multicolumn{1}{c}{} &
\multicolumn{1}{c}{(Mpc)} &
\multicolumn{1}{c}{L$_{FIR}$} &
\multicolumn{1}{c}{L$_{H\alpha}$} &
\multicolumn{1}{c}{Reference}
\\
\multicolumn{1}{c}{} &
\multicolumn{1}{c}{} &
\multicolumn{1}{c}{} &
\multicolumn{1}{c}{(L$_{\sun}$)} &
\multicolumn{1}{c}{$({\rm erg~s^{-1}})$}&
\multicolumn{1}{c}{} 
\\
}
\startdata

NGC 24  &  SAc &  7.8  &  8.2  &  40.0  &  SINGS  \\
NGC 337  &  SBd &  22.3  &  9.9  &  41.2  &  SINGS  \\
NGC 628  &  SAc &  9.9  &  9.6  &  41.3  &  SINGS  \\
NGC 925  &  SABd &  9.3  &  9.1  &  40.7  &  SINGS  \\
NGC 1097  &  SBb &  16.5  &  10.4  &  41.5  &  SINGS  \\
NGC 1291  &  SBa &  10.1  &  8.6  &  40.9  &  SINGS  \\
NGC 2403  &  SABcd &  4.6  &  9.2  &  41.2  &  Van Zee et al. (in prep.)  \\
NGC 2543  &  SBb &  37.4  &  9.9  &  41.2  &  Epinat et al. (2008)  \\
NGC 2639  &  SAa:? &  49.6  &  10.1  &  40.8  &  This work: WHT  \\
NGC 2841  &  SAb &  12.3  &  9.2  &  40.8  &  SINGS  \\
NGC 2857  &  SAc &  71.0  &  10.0  &  \ldots  &  \ldots  \\
NGC 3049  &  SBab &  24.1  &  9.4  &  40.9  &  SINGS  \\
NGC 3184  &  SABcd &  10.1  &  9.3  &  41.0  &  SINGS  \\
NGC 3344  &  SABbc &  6.9  &  9.0  &  40.6  &  Dale et al. (2009)  \\
NGC 3353  &  Sb?pec &  18.5  &  9.5  &  40.8  &  Hunter \& Elmegreen (2004)  \\
NGC 3367  &  SBc &  47.6  &  10.4  &  41.8  &  Garcia-Barreto \& Rosado (2001)  \\
NGC 3521  &  SABbc &  8.0  &  9.8  &  41.0  &  SINGS  \\
NGC 3621  &  Sad &  6.5  &  9.4  &  41.2  &  SINGS  \\
NGC 3633  &  SAa &  41.0  &  10.0  &  \ldots  &  \ldots  \\
NGC 3646  &  SAa &  65.2  &  \ldots  &  41.8  &  This work: WHT  \\
NGC 3938  &  SAc &  15.5  &  9.7  &  41.2  &  SINGS  \\
NGC 4254  &  SAc &  39.8  &  11.1  &  42.5  &  SINGS  \\
NGC 4321  &  SABbc &  14.1  &  10.0  &  41.0  &  SINGS  \\
NGC 4450  &  SAab &  14.1  &  8.9  &  40.0  &  SINGS  \\
NGC 4559  &  SABcd &  9.8  &  9.3  &  \ldots  &  \ldots  \\
NGC 4579  &  SABb &  13.9  &  9.4  &  40.9  &  SINGS  \\
NGC 4594  &  SAa &  12.7  &  9.2  &  \ldots  &  \ldots  \\
NGC 4725  &  SABab &  26.8  &  10.0  &  \ldots  &  \ldots  \\
NGC 4736  &  SAab &  4.8  &  9.4  &  40.7  &  Knapen et al. (2004)  \\
NGC 4826  &  SAab &  3.8  &  9.0  &  40.1  &  SINGS  \\
NGC 5055  &  SAbc &  8.3  &  9.8  &  41.1  &  SINGS  \\
NGC 5656  &  Sab &  51.4  &  10.2  &  \ldots  &  \ldots  \\
NGC 6373  &  SABc &  51.3  &  9.2  &  \ldots  &  \ldots  \\
NGC 6946  &  SABcd &  5.5  &  9.9  &  41.0  &  SINGS  \\
NGC 7331  &  SAb &  14.4  &  10.3  &  41.3  &  SINGS  \\
NGC 7793  &  SAd &  3.3  &  8.6  &  39.9  &  SINGS  \\
UGC 4704  &  Sdm: &  10.4  &  \ldots  &  \ldots  &  \ldots  \\
UGC 5853  &  Scd: &  132.6  &  10.0  &  \ldots  &  \ldots  \\
UGC 6879  &  SABd? &  37.3  &  8.9  &  \ldots  &  \ldots  \\
\enddata
{
$^{\dagger}$From the NASA Extragalactic Database (NED), using
H$_0$ = 73 km/s/Mpc, with Virgo,
Great Attractor, and Shapley Supercluster infall models.
$^{\ddagger}$Total 42.4 $-$ 122.5 $\mu$m far-infrared luminosity
from the Infrared
Astronomical Satellite (IRAS).
}
\end{deluxetable}
\clearpage